\documentclass[aps,pre,review,twocolumn,showpacs]{revtex4}
\usepackage{amsmath}
\usepackage{epsfig}
\usepackage{amssymb,graphicx}
\usepackage{hyperref}

\newfont{\logo}{logo10}

\newcommand{\bea}{\begin{eqnarray}}
\newcommand{\eea}{\end{eqnarray}}
\newcommand{\bes}{\begin{subequations}}
\newcommand{\ees}{\end{subequations}}
\newcommand{\ds}{\displaystyle}

\begin{document}
\bibliographystyle{revtex}

\title{General Multicomponent Yajima-Oikawa System: Painlev\'e Analysis, Soliton Solutions and Energy Sharing Collisions}
\author{T. Kanna\footnote{Corresponding author e-mail: kanna\_phy@bhc.edu.in}, K. Sakkaravarthi\footnote{ksakkaravarthi@gmail.com} and  K. Tamilselvan}
\affiliation{Post Graduate and Research Department of Physics, Bishop Heber College, Tiruchirappalli--620 017, Tamil Nadu, India}

\begin{abstract}
We consider the multicomponent Yajima-Oikawa (YO) system and show that the two-component YO system can be derived in a physical setting of a three-coupled nonlinear Schr\"odinger (3-CNLS) type system by the asymptotic reduction method. The derivation is further generalized to the multicomponent case. This set of equations describes the dynamics of nonlinear resonant interaction between a one-dimensional long wave and multiple short waves. The Painlev\'e analysis of the general multicomponent YO system shows that the underlying set of evolution equations is integrable for arbitrary nonlinearity coefficients which will result in three different sets of equations corresponding to positive, negative, and mixed nonlinearity coefficients. We obtain the general bright $N$-soliton solution of the multicomponent YO system in the Gram determinant form by using Hirota’s bilinearization method and explicitly analyze the one- and two-soliton solutions of the multicomponent YO system for the above mentioned three choices of nonlinearity coefficients.We also point out that the 3-CNLS system admits special asymptotic solitons of bright, dark, anti-dark, and gray types, when the long-wave--short-wave resonance takes place. The short-wave component solitons undergo two types of energy-sharing collisions. Specifically, in the two-component YO system, we demonstrate that two types of energy-sharing collisions---(i) energy switching with opposite nature for a particular soliton in two components and (ii) similar kind of energy switching for a given soliton in both components---result for two different choices of nonlinearity coefficients. The solitons appearing in the long-wave component always exhibit elastic collision whereas those of short-wave components exhibit standard elastic collisions only for a specific choice of parameters. We have also investigated the collision dynamics of asymptotic solitons in the original 3-CNLS system. For completeness, we explore the three-soliton interaction and demonstrate the pairwise nature of collisions and unravel the fascinating state restoration property.
\end{abstract}

\maketitle
\section{INTRODUCTION}
Nonlinear interactions of waves lead to a plethora of interesting physical phenomena in diverse areas of science that include nonlinear optics, water waves, plasma physics, condensed matter theory, and bio-physics \cite{whitham,scot,kivshar,dodd}. One such fascinating interaction process is long-wave--short-wave resonance interaction (LSRI), in which a resonant interaction takes place between a weakly dispersive long-wave and a short-wave packet when the phase velocity of the former (say $v_p$) exactly or almost matches the group velocity of the latter (say $v_g$), i.e., $v_p=v_g$, the Zakharov-Benny condition. Here the formation of the soliton in the short-wave (SW) component is due to a delicate balance of its dispersion by the nonlinear interaction of the long wave (LW) with the SW. Meanwhile, the formation and evolution of the soliton in the LW component is exclusively determined by the self-interaction of the short-wave packet. The theoretical formulation of this LSRI starts from the investigation of Zakharov \cite{zakharov} on Langmuir waves in plasma. Later, for unidirectional propagation of long waves, the general Zakharov equations were reduced to the Yajima-Oikawa (YO) equations describing one dimensional two-layer fluid flow \cite{oikawa}. In Ref. \cite{oikawa} itself, the YO system has been shown to be integrable by means of the sophisticated inverse scattering method and the soliton solutions have been obtained. In Ref. \cite{benny}, Benny has obtained independently the model equation describing the interaction between short wind-driven capillary gravity waves and gravity waves in deep water.

Following the above works on LSRI, there have been a flurry of research activities in the direction of LSRI involving a single SW component and a LW, in one dimension \cite{rede} as well as in two dimensions \cite{funakoshi}. Especially, in the context of nonlinear optics, the two-coupled nonlinear Schr\"odinger (CNLS) equations describing the interaction of two optical modes, one being in the normal group velocity dispersion (GVD) regime and the other being in the anomalous GVD regime, is shown to reduce to the single SW component YO system in the small-amplitude limit \cite{kivol}. Throughout this paper, we refer to the multicomponent YO system with $M$ number of SW components and a LW as $M$-YO system. Similar equations have also been shown to describe nonlinear three wave interaction of two nearly degenerate short waves with a long wave in photonic crystals \cite{3photonic}. Also, the LSRI has been investigated in various aspects in recent times \cite{recent,lsrinim,lsrirogue}. Particularly, in the negative refractive index media this LSRI has been investigated \cite{lsrinim}. The rogue waves of the one-component LSRI system have also been obtained recently  \cite{lsrirogue}.

Although there exists a significant amount of work for the integrable single-component YO-type system and its variants in 1D, results are scarce for their multicomponent counterparts that involve two or more SW components. Here, we briefly mention those interesting works. The energy transfer mechanism in protein molecules is explained by the description of the Zakharov system in Ref. \cite{davydov}. In Ref. \cite{ohtajpa7}, Ohta {\it et al} have derived the 2-SW LSRI system in (2+1) dimensions, where special soliton solutions of the underlying two-dimensional nonlinear evolutional equations have been obtained. The Painlev\'e integrability analysis of the ($2+1$)D LSRI system has been carried out in Ref. \cite{radha} and the exponentially localized (dromion) and periodic solutions were obtained using the truncated Painlev\'e approach \cite{radha}. The present authors (TK and KS) and their collaborators have obtained a more general $N$-soliton solution of the multicomponent (2+1)D LSRI system and revealed the fascinating energy sharing collision of bright solitons in the multicomponent two-dimensional LSRI system \cite{tklsri}. Also, the exact mixed (bright-dark) soliton solutions of the multicomponent (2+1)D LSRI system have been obtained and their underlying collisions were discussed in detail \cite{tklsribd}.

Coming back to the one-dimensional case, the two-component YO-type system is also  obtained in the study of interaction of quasiresonant two-frequency short pulses with a long wave \cite{sazonov} and for the case of high anisotropy the underlying system is shown to be integrable and soliton solutions are obtained. Recent studies also indicate that the multicomponent YO system will find promising applications in condensed matter theory, particularly in spinor condensates \cite{jetp2009}. More interestingly, in a classic work \cite{myrza86}, multicomponent YO-type equations have been derived in the context of the many-component magnon-phonon system and its corresponding Hamiltonian formalism was developed, but the complete dynamics of solitons in such system still remains unexplored. In fact, even for the two-component YO system the propagation and interaction dynamics have to be completely unraveled.

In what follows, we investigate the dynamics of general multicomponent YO equations. To bring out the physical importance of the  general multicomponent ($1+1$)-dimensional YO system, first we derive 2-component YO system from the 3-CNLS equations, governing the dynamics of multiple-pulse propagation in Kerr-type nonlinear media. For this purpose, we make use of the asymptotic reduction method \cite{frantz,frantz2,frantz3}, an appropriate method for dealing with multiple time scales. Such multiple-scale method is widely used in the literature to reduce a general nonlinear evolution equation to an asymptotic equation which is integrable by means of appropriate transformation of variables. Thus, this method provides an easy way to investigate the dynamics of the original nonlinear system even though it is non-integrable, with the help of asymptotic solutions of the reduced system. Very recently, in Ref. \cite{frantz}, the (2+1) dimensional CNLS system governing the dynamics of a binary mixture of disk-shaped Bose-Einstein condensates is reduced to the integrable (2+1)D LSRI system by applying the asymptotic reduction procedure. In nonresonant quadratic nonlinear media, by applying this method several single-component and multicomponent asymptotic evolution equations have been obtained and their solutions are analyzed in detail \cite{ablopre2001}.

Then, we focus our attention on the integrability aspects of the physically relevant general $M$-YO  system at hand by performing the Painlev\'e singularity structure analysis  \cite{radha,ablobook,weiss,painml}, a useful analytical tool to isolate and identify integrable dynamical system in a compact way. A dynamical system (which is described by a nonlinear partial differential equation) is said to be Painlev\'e integrable if its solutions are single valued in the neighborhood of a movable singularity manifold and exhibit sufficient number of integrals of motion \cite{radha,ablobook,weiss,painml}. We construct the exact bright soliton solutions of the obtained integrable version of the system by using Hirota's bilinearization method \cite{hirota}, one of the powerful analytical tools to construct the soliton solution of integrable equations due to its algebraic nature.

The organization of the paper is as follows. In Sec. \ref{secmodel}, we explicitly derive the ($1+1$)D 2-YO system from a setting of the 3-CNLS system by using the asymptotic reduction method and generalize the results to the multicomponent case. The integrability nature of the obtained system is explored by using the Painlev\'e singularity structure analysis in Sec. \ref{secpain}. We present the bright multisoliton solution of the integrable $M$-YO system using Hirota's direct method and study their dynamics in Sec. \ref{secsoli}. The soliton collision dynamics is presented in Sec. \ref{collisions} and the results are summarized in the final section.

\section{The Description of The Model}\label{secmodel}
The nonlinear resonance interaction between multiple (say $M$) short waves and a long wave can be described by the following nondimensional multicomponent generalization of the Yajima-Oikawa system
\bes\bea
&&i S_{t}^{(\ell)}+ S_{xx}^{(\ell)}+ L S^{(\ell)}=0, \qquad \ell=1,2,3,...,M, \\
&&L_{t}=\sum_{\ell=1}^M c_{\ell} |S^{(\ell)}|_x^2,\eea\label{model}\ees
where $S^{(\ell)}$ and $L$, respectively, indicate the $\ell$-th short-wave and long-wave components. In Eq. (\ref{model}), `$t$' and `$x$' represent the partial derivatives with respect to evolutional and spatial coordinates, respectively, and the nonlinearity coefficients $c_{\ell}$ are arbitrary real parameters.

To emphasize the physical significance of the above system (\ref{model}), in the following we demonstrate that the general two-component YO system, with $M=2$ and arbitrary $c_1$ and $c_2$, follows from a general setting of 3-CNLS equations, by applying the asymptotic reduction procedure. For $M=2$, Eq. (\ref{model}) describes the resonance interaction of two SWs with a LW. Hence, in order to deduce the corresponding 2-YO equations we consider the propagation of three optical fields in a triple mode optical fiber. The interaction of three optical modes in such nonlinear optical fiber, with one pulse in the normal dispersion regime and the remaining pulses in the anomalous dispersion regime, is governed by the following adimensional three-component NLS-type equations:
\bes\bea
 i q_{1,Z}-\frac{1}{2}q_{1,TT} + (\sigma_{11} |q_1|^2 + \sigma_{12} |q_2|^2 + \sigma_{13}|q_3|^2) q_1=0,~\\
 i q_{2,Z}+\frac{1}{2}q_{2,TT} + (\sigma_{21} |q_1|^2 + \sigma_{22} |q_2|^2 + \sigma_{23}|q_3|^2) q_2=0,~\\
 i q_{3,Z}+\frac{1}{2}q_{3,TT} + (\sigma_{31} |q_1|^2 + \sigma_{32} |q_2|^2 + \sigma_{33}|q_3|^2) q_3=0,~
\eea\label{eq1}\ees
where $q_j,~j=1,2,3,$ represents the propagation mode, the subscripts $Z$ and $T$ indicate the propagation direction and retarded time, respectively, and $\sigma_{ij}$ ($i,j=1,2,3$) are self-phase modulation (SPM) coefficients and cross-phase modulation (XPM) coefficients for $i=j$ and $i\neq j$, respectively. System (\ref{eq1}) is a generalization of the two-component case considered in Ref. \cite{kivol}. Various multicomponent generalizations  of Eq. (\ref{eq1}) also appear in several different contexts as mentioned in the introduction, notably, in nonlinear optics, plasma physics, hydrodynamics, Bose-Einstein condensates, and bio-physics. The recent developments in artificial metamaterials and photonic crystal fibers make it feasible to achieve the negative values also for the nonlinearity coefficients $\sigma_{ij}$. Also, a similar type of CNLS equations have been investigated recently \cite{bosefermi} to describe a quantum degenerate mixture of bosons and fermions. It is instructive to point out that system (\ref{eq1}) is integrable for the choice \cite{ablobook,makhankov1981,tkpla,rkml,zakh}
\bes\bea
\sigma_{j1}=\sigma_1, \quad \sigma_{j2}=\sigma_2, \quad \sigma_{j3}=\sigma_3, \\
\sigma_{1\ell}=-\sigma_{\ell}, \quad j=2,3, \quad \ell=1,2,3,
\eea\label{intcon}\ees
and admits different types of solitons namely, bright, dark, bright-dark/dark-bright type solitons.

We apply the Madelung transformation $q_1=\sqrt{\rho}~e^{i\theta}$, where the real valued functions $\rho(Z,T)$ and $\theta(Z,T)$ denote normalized density and phase, respectively, to Eq. (\ref{eq1}). Then the real and imaginary parts are grouped together separately, to obtain the following system of hydrodynamic equations:
\bes\bea
&\theta_Z-\frac{(\theta_T)^2}{2}+\frac{(\sqrt{\rho})_{TT}}{2\sqrt{\rho}}-\sigma_{11}{\rho}-\sigma_{12}|q_2|^2-\sigma_{13}|q_3|^2=0,&~~~\\
&\rho_Z-(\rho\theta_T)_T=0,~~~~~~~~~~~~~~~~~~~~~~~~~~~~~~~~~~~~~~~~~~~~~~~&\\
&iq_{2,Z}+\frac{1}{2}q_{2,TT}+\sigma_{21}{\rho q_2}+(\sigma_{22}|q_2|^2+\sigma_{23}|q_3|^2){q_2}=0,&~\\
&iq_{3,Z}+\frac{1}{2}q_{3,TT}+\sigma_{31}{\rho q_3}+(\sigma_{32}|q_2|^2+\sigma_{33}|q_3|^2){q_3}=0.&~
\eea\label{eq3}\ees
Next, we expand $\rho$ and $\theta$ around the trivial solutions $\rho=1$ and $\theta=\sigma_{11}Z$, resulting in the absence of $q_2$ and $q_3$, as $\rho=1+\varepsilon{r(Z,T)}$, $\theta=\sigma_{11}Z+{\varepsilon}{\phi(Z,T)}$, with $\varepsilon$ being a formal small expansion parameter. By expressing $q_2$ and $q_3$ as $q_2=\varepsilon{s_1}e^{i\sigma_{21} Z}$ and $q_3=\varepsilon{s_2}e^{i\sigma_{31} Z}$ in Eqs. (\ref{eq3}a) and (\ref{eq3}b) and collecting the terms arising at the lowest power of $\varepsilon$, we obtain the following set of equations:
\bes\bea
&& r_{Z}-\phi_{TT} =0, \\  &&\phi_Z+\frac{1}{4}r_{TT}-\sigma_{11}{r}=0.
\eea\label{eq6}\ees
Equations (\ref{eq6}) can be combined together into a single equation as $r_{ZZ}-\sigma_{11}r_{TT}+\frac{1}{4}{r_{TTTT}}=0$ and from the dispersion relation of this linear equation, we find that in the long-wavelength limit the small-amplitude wave travels with a velocity $c=\sqrt{\sigma_{11}}$, i.e., the phase velocity of the long wave. This requires that $\sigma_{11}$ be positive. Otherwise, the velocity will become complex and there will be absorption.

In order to obtain the two-component YO system, we rewrite Eq. (\ref{eq3}) in terms of new variables `$x$' and `$t$', and consider a frame moving with velocity $c~(=\sqrt{\sigma_{11}})$ through a scaling transformation $x=\sqrt{\varepsilon}(T-\sqrt{\sigma_{11}}Z)$ and $t=\varepsilon{Z}$. By expressing $\rho$, $\theta$, $q_2$ and $q_3$ as
\bes\bea
&&\rho=1+\varepsilon{r(x,t)},\\
&&\theta=\sigma_{11}Z+\sqrt{\varepsilon}{\phi(x,t)},\\
&&q_2=\varepsilon^{3/4}{v(x,t)}e^{i(k T-(\omega-\sigma_{21}) Z)},\\
&&q_3=\varepsilon^{3/4}{w(x,t)}e^{i(k T-(\omega-\sigma_{31}) Z)},\eea\label{trans}\ees
where $k$ and $\omega$ are wave number and frequency, respectively, we get the self-consistent equations $\phi_x=-\sqrt{\sigma_{11}}~{r}$ and $\phi_t=\sigma_{12}|v|^2+\sigma_{13}|w|^2$, from Eq. (\ref{eq3}a), respectively at the orders $\varepsilon$ and at $\varepsilon^{3/2}$. Then the evolution of `$r$' is deduced as
\bes\bea
r_t&=&-\left(\frac{\sigma_{12}}{\sqrt{\sigma_{11}}}|v|^2_{x}+\frac{\sigma_{13}}{\sqrt{\sigma_{11}}}|w|^2_{x}\right).
\eea
Similarly, from Eqs. (\ref{eq3}c) and (\ref{eq3}d) we obtain the following equations by making use of Eq.(\ref{trans})
\bea
i v_{t}+\frac{1}{2} v_{xx}+\sigma_{21}r\upsilon =0,\\
 iw_{t}+\frac{1}{2} w_{xx}+\sigma_{31}r\omega =0,
\eea\label{re2}\ees
with a dispersion relation $\omega=k^2/2$, where $k=\sqrt{\sigma_{11}}$. From this we find that the group velocity of short waves ($\frac{d\omega}{dk}$) is equal to the phase velocity of the long wave ($c=\sqrt{\sigma_{11}}$). This is nothing but the Zakharov-Benny condition for resonance interaction between the long wave and two short waves. A careful comparison of the reduced system (\ref{re2}) with the original system Eq. (\ref{eq1}) shows that, when the LSRI occurs, the XPM contributions of the $q_1$ component to the $q_2$ and $q_3$ components and also their XPM contribution to $q_1$ along with its own SPM only play a pivotal role in determining the nature of such nonlinear interactions. This kind of realization is indeed possible in atomic systems describing the propagation of intense optical beams of different frequencies in a cascaded configuration \cite{cascaded}. Also, one can notice another important point that, when LSRI takes place, apart from the choice (\ref{intcon}) system (\ref{eq1}) can admit soliton-like structures for another choice $\sigma_{21}=\sigma_{31}\equiv \sigma$ and can be different from $\sigma_{11}$, with $\sigma_{12}$ and $\sigma_{13}$ being arbitrary, as will be shown in Sec. \ref{secsoli}. In fact, Eq. (\ref{re2}) can be reexpressed as a two-component Yajima-Oikawa system,
\bes\bea
&&i S_{t}^{(1)}+S_{xx}^{(1)}+ L S^{(1)}=0,\\
&&i S_{t}^{(2)}+S_{xx}^{(2)}+ L S^{(2)}=0,\\
&&L_{t}=c_1|S^{(1)}|_x^2+c_2|S^{(2)}|_x^2,\\
\mbox{where} &&c_1=-\sqrt{2/\sigma_{11}}~\sigma_{12}\sigma,~ c_2=-\sqrt{2/\sigma_{11}}~\sigma_{13}\sigma,~~~ \label{choice}
\eea \label{model2c}\ees
with the choice $\sigma_{21}=\sigma_{31}=\sigma$ and making use of the transformations $\sigma r \rightarrow L$, $v\rightarrow S^{(1)}$, $w \rightarrow S^{(2)}$ and $x \rightarrow \sqrt{2}x$. The above set of evolution equations (\ref{model2c}) with $c_1=c_2=1$ arises in the description of the quasiresonant two-frequency pulse propagation with a long electromagnetic wave generated in an asymmetric medium \cite{sazonov}, and also in the study of Langmuir waves, in the Davydov model with two excitonic modes coupled with a phonon field \cite{davydov}. In Eq. (\ref{model2c}), though the absolute values of $c_1$ and $c_2$ can be absorbed, the solutions will behave differently if their signs differ. In order to deal with such general equations we retain $c_1$ and $c_2$ explicitly in Eq. (\ref{model2c}).

Recently, multicomponent CNLS equations have been derived as evolution equations in a physical set up, particularly for multiple pulse propagation in multi-mode optical fibers \cite{meccozi}. A straightforward application of the above discussed asymptotic reduction procedure to such $m$-component CNLS setting, with arbitrary $m$, describing the interaction of $m$ optical modes will result in the $M$ ($\equiv m-1$) component YO system (\ref{model})  with the nonlinearity coefficients $c_{\ell}$ ($=-\sqrt{2/\sigma_{11}}~\sigma_{1\ell+1}\sigma$, $\ell=1,2,3,...,M$). Moreover, the system (\ref{model}) with nonlinearity coefficient $c_{\ell}=+1$ for $\ell=1,2,...,p$, and $c_{\ell}=-1$ for  $\ell=p+1, p+2,...,M$, arises as the governing equation in a magnetic system which describes the magnon-phonon interactions of a many-sublattice isotropic XY chain at the long-wavelength limit \cite{myrza86}, as mentioned in the introduction.

In order to reveal the importance of the choice of nonlinearity coefficients in the $M$-YO  system (\ref{model}) and due to its diversified applications, in this paper we focus our attention on studying the integrable nature of system (\ref{model}) and the underlying soliton solutions. First, in the following section, we show that Eq. (\ref{model}) is integrable by performing the Painlev\'e analysis. Then we construct the multisoliton solutions corresponding to the identified integrable choices and explore their collision dynamics.

\section{Painlev\'e Integrability Analysis}\label{secpain}
Here, we apply the Painlev\'e test to study the integrability nature of the $M$-YO  system (\ref{model}). We rewrite Eqs. (\ref{model}) and the complex conjugate equation of (\ref{model}a) in terms of new dependent variables ($S^{(\ell)}=m^{(\ell)},~S^{(\ell)*}=n^{(\ell)}$ and $L=l$) as
\bes\bea
&& im^{(\ell)}_t+  m^{(\ell)}_{xx}+  l m^{(\ell)}=0, \quad \ell=1,2,3,...,M,\\
&& -in_t^{(\ell)}+  n_{xx}^{(\ell)}+  l n^{(\ell)}=0,\quad \ell=1,2,3,...,M,\\
&& l_{t}=\sum_{\ell=1}^M (c_{\ell} m^{(\ell)}n^{(\ell)})_x.
\eea\label{pa2}\ees
The Painlev\'e analysis is carried out by looking for a generalized Laurent expansion for the dependent variables
\bea
(m^{(\ell)},n^{(\ell)},l)=\left(\sum_{j=0}^n m_j^{(\ell)} \phi^{j+\alpha_{\ell}},\sum_{j=0}^n n_j^{(\ell)} \phi^{j+\beta_{\ell}},\sum_{j=0}^n l_j \phi^{j+\gamma}\right)\hspace{-0.15cm},~\label{ls}
\eea
in the neighborhood of the noncharacteristic singular manifold $\phi(x,t)$, with nonvanishing derivatives with respect to `$x$' and `$t$'; i.e., $\phi_t \neq 0$ and $\phi_x \neq 0$. Here $m_j^{(\ell)},~n_j^{(\ell)}$, and $l_{j}$ are arbitrary analytic functions of $x$ and $t$, while $\alpha_{\ell},~\beta_{\ell}$, and $\gamma$ are integers to be determined.\\

\noindent \emph{Leading order analysis}: As a first step of the Painlev\'e test, we perform a leading order analysis by assuming the forms of the dependent variables [by terminating the Laurent series (\ref{ls})] as $m^{(\ell)}=m_{0}^{(\ell)}\phi^{\alpha_{\ell}},\quad n=n_{0}^{(\ell)}\phi^{\beta_{\ell}}$, and $l=l_{0}\phi^{\gamma}$. Substituting these functions in Eq. (\ref{pa2}), and balancing the most dominant powers of $\phi$, we find $\alpha_{\ell}=\beta_{\ell}=-1$ and $\gamma=-2$ and obtain the corresponding leading order equations, arising at the order of $\phi^{-3}$, as
\bes\bea
&& l_0=-2 \phi_x^2, \qquad \mbox{($2M$ times)}\\
&& -2\phi_x\phi_t=\sum_{\ell=1}^M c_{\ell} m_0^{(\ell)}n_0^{(\ell)}.
\eea\label{mloe2}\ees
\emph{Resonances}: The standard next step is to identify the resonances (powers) at which arbitrary functions can enter into the Laurent series (\ref{ls}). By substituting the general Laurent series expansion (\ref{ls}) for $\alpha_{\ell}=\beta_{\ell}=-1$ and $\gamma=-2$ into Eq. (\ref{pa2}), we obtain a set of equations at the order of $\phi^{j-3}$, which are expressed in a  block form as
\bea
\left(\begin{array}{cc}
\mathbb{A} & \mathbb{P} \\
\mathbb{B}&(j-2)\phi_{t}\\
\end{array}\right) \mathbb{Q}={\bf 0}, \label{resmat}
\eea
where the block matrices $\mathbb{A}$, $\mathbb{B}$, $\mathbb{P}$ and $\mathbb{Q}$ of dimension ($2M\times 2M$), ($1\times 2M$), ($2M\times 1$) and ($(2M+1)\times 1$), respectively, are defined as $\mathbb{A}=j(j-3)\phi_{x}^2 \mathbb{I}$, $\mathbb{B}=(j-2)\phi_{x}(c_1n_0^{(1)}, c_1m_0^{(1)}, c_2n_0^{(2)}, c_2m_0^{(2)},...,c_{M} n_0^{(M)},~c_{M} m_0^{(M)})$, $\mathbb{P}=(m_0^{(1)}, n_0^{(1)}, m_0^{(2)}, n_0^{(2)},..., m_0^{(M)}, n_0^{(M)})^T$ and $\mathbb{Q}=(m_j^{(1)}, n_j^{(1)}, m_j^{(2)}, n_j^{(2)},..., m_j^{(M)}, n_j^{(M)}, l_j)^T$. Here, $\mathbb{I}$ is the ($2M\times 2M$) identity matrix and `$T$' appearing in the superscript represents the transpose of the matrix. One can obtain the following ($4M+1$) number of integer resonances:
\bea
j=-1,\overbrace{0,\cdots,~0,}^{(2M-1)}~2,\overbrace{3,\cdots,~3,}^{(2M-1)}~4.
\eea
\emph{Arbitrary analysis}: As a final step, one has to prove the arbitrariness of each resonance. Obviously, the resonance $j=-1$ corresponds to the arbitrariness of non-characteristic manifold $\phi(x,t)$. One can also prove the existence of a sufficient number of arbitrary parameters at other resonance values as well.

In order to have a clear picture of the above Painlev\'e analysis, we explicitly give the results for the three-component YO system [i.e., Eq. (\ref{model}] with $M=3$). In this case, we get the leading order equations as
\bes\bea
 l_0&=&-2 \phi_x^2, \qquad \mbox{(6 times)}\\
 -2\phi_x\phi_t&=&c_{1} m_0^{(1)}n_0^{(1)}+c_{2} m_0^{(2)}n_0^{(2)}+c_{3} m_0^{(3)}n_0^{(3)},~~~~
\eea\label{loe2}\ees
and the explicit form of resonance equation is given below:
\bea
&& j^5[\Omega j^8-20\Omega j^7+(171\Omega +2\Delta)j^6-(810\Omega +34\Delta)j^5 \nonumber\\
&& ~~~~~~~+(2295\Omega +240\Delta)j^4 -(3888\Omega +900\Delta)j^3 \nonumber\\
&& ~~~~~~~+(3645\Omega +1890\Delta)j^2 -(1458\Omega +2106\Delta)j\nonumber\\
&& ~~~~~~~+972\Delta]=0,~~~~~\label{res}
\eea
where $\Omega= \phi_t \phi_x^{12}$ and $\Delta=(c_1 m_0^{(1)}n_0^{(1)}+ c_2 p_0^{(2)}q_0^{(2)}+ c_3 r_0^{(3)}s_0^{(3)})\phi_x^{11}$. By making use of the leading order equation (\ref{loe2}b), the roots of Eq. (\ref{res}), i.e., the resonances, are found to be
\bea
j=-1,~0,~0,~0,~0,~0,~2,~3,~3,~3,~3,~3,~4.
\eea
Here, among the thirteen resonances, the negative resonance $j=-1$ corresponds to the arbitrariness of the manifold $\phi$. We have performed the arbitrary analysis indicating the existence of a sufficient number of arbitrary functions at each resonance in Appendix A. Thus we can conclude that the three-component YO system passes the Painlev\'e test and is integrable in the Painlev\'e sense.

As in the case of the three-component system, one can also explicitly prove the existence of a sufficient number of arbitrary parameters at each resonance value with the help of symbolic computation for an arbitrary $M$-component case. Thus the above analysis clearly demonstrates that the $M$-YO  system (\ref{model}) is integrable by means of Painlev\'e singularity structure analysis, for arbitrary $c_{\ell},~\ell=1,2,3,...,M$, which can admit both positive and negative real values. Note that the signs of $c_{\ell}$ can alter the dynamics dramatically as will be shown in the following sections. To the best of our knowledge, such type of study has not been carried out so far for the general multicomponent Yajima-Oikawa system (\ref{model}).

\section{Bright soliton solutions}\label{secsoli}
In this section, we obtain the bright $N$-soliton solution of multicomponent YO system (\ref{model}), with arbitrary $N$, by employing the Hirota  bilinearization method \cite{hirota}, a well designed analytical technique of algebraic nature known for its efficiency to generalize the results for arbitrary $N$-soliton solutions. For this purpose, first we bilinearize (\ref{model}) with the bilinearizing  transformations $S^{(\ell)}=\frac{g^{(\ell)}}{f}$, $\ell=1,2,...M$, and $L=2\frac{\partial^2}{\partial x^2}(\ln{f})$, where $g^{(\ell)}$ are complex functions and $f$ is a real function. Then in bilinear form Eq. (\ref{model}) can be expressed as
\bes\bea
&&(iD_t+D_x^2)g^{(\ell)} f=0,\quad \quad \ell=1,2,...M,\\
&&D_tD_x (f f)=\sum_{\ell=1}^{M} c_{\ell} g^{(\ell)}g^{(\ell)*},
\eea\label{beq}\ees
where $*$ appearing in the superscript denotes the complex conjugation, $D_x$ and $D_t$ are the standard Hirota bilinear operators \cite{hirota}, $D_x^{p}D_t^{q}(a\cdot b) =\big(\frac{\partial}{\partial x}-\frac{\partial}{\partial x'}\big)^p\big(\frac{\partial}{\partial t}-\frac{\partial}{\partial t'}\big)^q a(x,t)b(x',t')\big|_{ \ds (x=x', t=t')}$. Expanding $g^{(\ell)}$'s and $f$ in formal power series in terms of a small arbitrary real parameter and following the standard procedure \cite{hirota,rkpre,tkprl,tkpre,tkjpa} one can obtain exact multi-soliton solutions.

\subsection{Bright $N$-soliton solution of $M$-YO system (\ref{model})}
The explicit form of the bright $N$-soliton solution, for arbitrary $N$, is obtained by expressing the power series expansion for the dependent variables $g^{(\ell)}$ and $f$ as $g^{(\ell)}=\chi g_1^{(\ell)}+\chi^3 g_3^{(\ell)}+ \cdots +\chi^{(2N-1)} g_{(2N-1)}^{(\ell)}$ and $f=1+\chi^2 f_2+\chi^4 f_4+\cdots + \chi^{2N} f_{2N}$, respectively, and by recursively solving the resultant set of equations arising at different powers of $\chi$. We express below the obtained bright $N$-soliton solution in Gram determinant form \cite{hirota} and show that indeed the $N$-soliton solution satisfies the bilinear equations (\ref{beq}).

The general bright $N$-soliton solution of the $M$-YO system can be written as
\bes\bea
S^{(\ell)}&=&\frac{g^{(\ell)}}{f}, \quad \ell=1, 2, \ldots, M,\\
L&=&2\frac{\partial^2}{\partial x^2}(\ln{f}),
\eea
where
\bea
g^{(\ell)}=
\left|
\begin{array}{ccc}
A & I & \phi\\
-I & B & {\bf 0}^T\\
{\bf 0} & a_{\ell} & 0
\end{array}
\right|, \quad \quad f= \left|
\begin{array}{cc}
A & I\\
-I & B
\end{array}
\right|.
\eea
Here $A$ is a ($N \times N$) Gramian with elements $A_{pj}= \frac{e^{\eta_{p}+\eta_{j}}}{(k_{p}+k_{j}^*)}$, $I$ is an identity matrix of order ($N \times N$), $\bf{0}$ is a ($1 \times N$) null matrix and $B$ is a ($N \times N$) constant block matrix whose elements are defined as
\bea
B_{pj}=\kappa_{jp}=\frac{\psi_p^{\dagger}c\psi_j}{2i(k_j^2+k_p^{*2})} \equiv \frac{\ds\sum_{\ell=1}^M c_{\ell}\alpha_p^{(\ell)*}\alpha_j^{(\ell)}}{2i(k_j^2-k_p^{*2})},~ 
\eea
where $p,j=1, 2,...,N$. Also, $a_{\ell}$ is a row matrix of order ($1\times M$), $\psi_j$ and $\phi$ are column matrices of order ($M \times 1$) and  ($N \times 1$), respectively, and $c$ is a ($M \times M$) diagonal matrix defined as
\bea
a_{\ell} &=& -\left(\alpha_1^{(\ell)}, \alpha_2^{(\ell)}, \ldots, \alpha_{N}^{(\ell)}\right), \quad \ell=1,2,3,...,M, \qquad \\ \psi_j&=&\left(\alpha_j^{(1)},~\alpha_j^{(2)},\ldots,\alpha_j^{(M)}\right)^T, \quad j=1, 2, \ldots, N,\\
\phi &=& \left(e^{\eta_1},e^{\eta_2},\ldots,e^{\eta_{N}}\right)^T,~ c= \mbox{diag}\left(c_1, c_2, \ldots, c_M \right).~~~~~~~
\eea\label{nsol}\ees
In Eq. (\ref{nsol}), $M$ and $N$ represent the component number and soliton number, respectively, and $\eta_j=k_j(x+i k_j t)$, where $k_j$ and $\alpha_{j}^{(\ell)}$, $\ell=1,2,\ldots,M$, $j=1,2,\ldots,N$, are arbitrary complex parameters. Here the symbols $\dagger$ and $T$ appearing in the superscript indicate the transpose conjugate and transpose of the matrix, respectively.

\subsection{Proof of $N$-soliton solution}
In order to prove that the $N$-soliton solution (\ref{nsol}) satisfies the bilinear equations (\ref{beq}), we find the derivatives of the above mentioned determinants $g^{(\ell)}$ and $f$ by using the derivative identities of determinants in a standard form \cite{hirota,tklsri}  and present them in Appendix B. By substituting the expressions for $g^{(\ell)}$, $g^{(\ell)}_{t}$, $g^{(\ell)}_x$, $g^{(\ell)}_{xx}$, $f$, $f_t$, $f_x$, and  $f_{xx}$, in the first bilinear equation (\ref{beq}a), we arrive at the following relation which is nothing but the Jacobian identity \cite{hirota}:
\bes\bea
&&\left|
\begin{array}{cccc}
A & I & \phi &\phi_{x}\\
-I & B & {\bf 0}^T& {\bf 0}^T\\
{\bf 0} & a_{\ell} & 0 &0\\
-\phi^{\dagger}& {\bf 0}&0 &0
\end{array}
\right|
\left|
\begin{array}{cc}
A & I\\
-I & B
\end{array}
\right| \nonumber\\
&&~~=
\left|
\begin{array}{ccc}
A & I & \phi_x \\
-I & B & {\bf 0}^T\\
-\phi^{\dagger} & {\bf 0} & 0
\end{array}
\right|
\left|
\begin{array}{ccc}
A & I & \phi\\
-I & B & {\bf 0}^T\\
{\bf 0} & a_{\ell} & 0
\end{array}
\right| \nonumber\\
&&~~~~ -
\left|
\begin{array}{ccc}
A & I &  \phi_x \\
-I & B  & {\bf 0}^T\\
{\bf 0} & a_{\ell} & 0
\end{array}
\right|
 \left|
\begin{array}{ccc}
A & I & \phi \\
-I & B & {\bf 0}^T\\
-\phi^\dagger & {\bf 0} & 0
\end{array}
\right|.~~~~~~~~
\eea
Similarly, by substituting the expression for $g^{(\ell)}$, $g^{(\ell)*}$, $f$, $f_x$, $f_t$ and $f_{xt}$ in the second bilinear equation (\ref{beq}b), we find that Eq. (\ref{beq}b) too can be expressed as a Jacobian identity.
\bea
&&\sum_{\ell=1}^M c_{\ell}
\left|
\begin{array}{cccc}
A & I & \phi &{\bf 0}^T \\
-I & B & {\bf 0}^T& -a_{\ell}^{\dagger}\\
-\phi^{\dagger}&{\bf 0} & 0 & 0 \\
{\bf 0}& a_{\ell}&0 &0
\end{array}
\right|
\left|
\begin{array}{cc}
A & I\\
-I & B
\end{array}
\right|\nonumber\\
&&~~ =
\sum_{\ell=1}^M c_{\ell}
\left|
\begin{array}{ccc}
A & I & {\bf 0}^T \\
-I & B & -a_{\ell}^{\dagger}\\
{\bf 0} & a_{\ell} & 0
\end{array}
\right|
\left|
\begin{array}{ccc}
A & I & \phi \\
-I & B & {\bf 0}^T\\
-\phi^\dagger & {\bf 0} & 0
\end{array}
\right| \nonumber\\
&&~~~~ -\sum_{\ell=1}^M c_{\ell}
\left|
\begin{array}{ccc}
A & I & {\bf 0}^T\\
-I & B & -a_{\ell}^{\dagger}\\
-\phi^\dagger & {\bf 0} & 0
\end{array}
\right|
\left|
\begin{array}{ccc}
A & I & \phi\\
-I & B & {\bf 0}^T\\
{\bf 0} & a_{\ell} & 0
\end{array}
\right|.
\eea \label{msolproof}\ees
Thus the bright $N$-soliton solution of $M$-YO system (\ref{model}), expressed in the Gram determinant form (\ref{nsol}) indeed satisfies the bilinear equations (\ref{beq}). However, to get insight into the real physics behind this solution one has to write down the explicit form of the solution and analyze the resulting expressions in detail.

\subsection{Bright one-soliton solution}
In its explicit form the bright one-soliton solution of the $M$-YO system (\ref{model}) (Eq. (\ref{nsol}) with $N=1$) reads as
\bes\bea
&&S^{(\ell)}=  2A_{\ell}k_{1R} \sqrt{k_{1I}}~\mbox{sech} \left(\eta_{1R}+\frac{R}{2}\right) e^{i(\eta_{1I}+\frac{\pi}{2})},~~~~~\\
&&L=2k_{1R}^2 \mbox{sech}^2 \left(\eta_{1R}+\frac{R}{2}\right), \quad \ell=1,2,...M,
\eea\label{1sol}\ees
where $A_{\ell}={\alpha_1^{(\ell)}}{\left(\ds\sum_{\ell=1}^{M}c_{\ell} |\alpha_1^{(\ell)}|^2\right)^{-\frac{1}{2}}}$, $e^R=\frac{-\ds\sum_{\ell=1}^{M} c_{\ell} |\alpha_1^{(\ell)}|^2}{16k_{1R}^2k_{1I}}$, $\eta_{1R}=k_{1R}(x+2k_{1I}t)$ and $\eta_{1I}=k_{1I}x+(k_{1R}^2-k_{1I}^2)t$. Here and in the following, $R$ and $I$ appearing in the subscript represent the real and imaginary parts of a given parameter, respectively. The above one-soliton solution of the $M$-YO system is characterized by ($M+1$) arbitrary complex parameters, $\alpha_1^{(\ell)},~\ell=1,2,...,M,$ and $k_1$. The $A_{\ell}$'s defined below (\ref{1sol}) can be ascribed to the polarization of the SW components.  Here the nonlinearity coefficients ($c_{\ell}$) can take any arbitrary real values (both positive and negative). The amplitudes of the propagating soliton in the short- and long-wave components are $2A_{\ell}k_{1R} \sqrt{k_{1I}}$ and $2k_{1R}^2$, respectively, and their velocity is $-2k_{1I}$. The interesting feature of this solution is that the explicit appearance of soliton velocity in the amplitude part of the SW component. As a consequence of this, the taller soliton will travel faster, a behavior akin to Korteweg-deVries (KdV) solitons. It is noteworthy to remark that such velocity dependent amplitude is not at all possible in the propagation of solitons in Manakov type (CNLS) systems. Here, such dependence of amplitude on velocity is a signature of the long-wave--short-wave resonance interaction. Additionally, though the form of bright solitons in long-wave component looks like KdV soliton, its amplitude is independent of the velocity and $\alpha_1^{(\ell)}$ parameters and vice-versa. Thus by tuning the nonlinearity coefficients one can profitably adjust the nature of solitons appearing in the SW components without altering the soliton appearing in the LW component.

The above one-soliton solution will feature both singular and nonsingular (regular) soliton solutions depending on the parameters $\alpha_1^{(\ell)}$, $c_{\ell}$ and $k_{1I}$ or simply the value of $e^R$. Based on the nature of nonlinearity coefficients $c_{\ell}$, the one-soliton solution of $M$-YO system (\ref{model}) can be classified into the following three cases:\\
\indent (i) positive nonlinearity coefficients ($c_{\ell}>0$),\\
\indent (ii) negative nonlinearity coefficients ($c_{\ell}<0)$,\\
\indent (iii) mixed-type coefficients (both positive and negative values of $c_{\ell}$).\\

\begin{figure}[h]
\centering\includegraphics[width=1.0\linewidth]{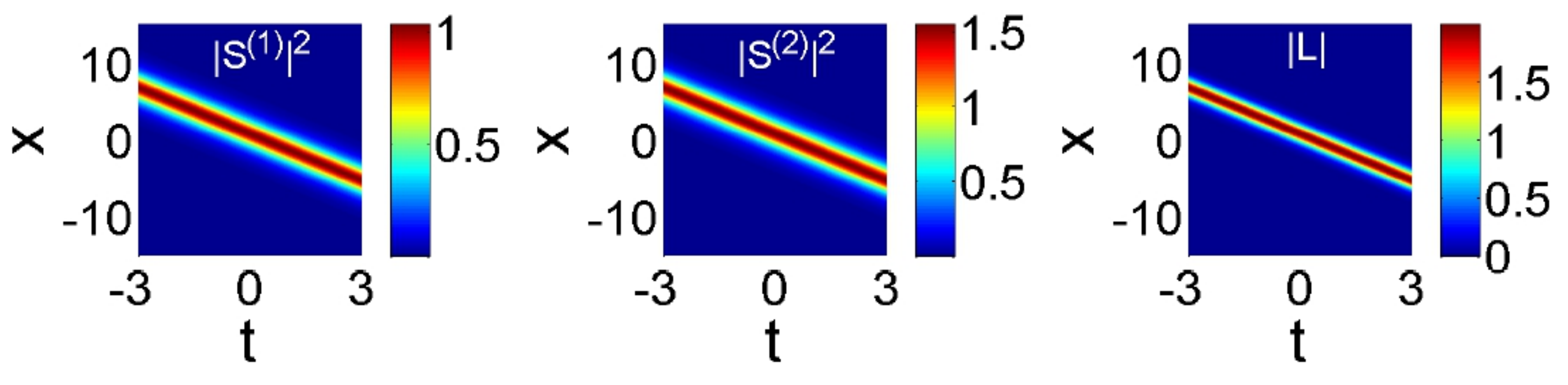}
\caption{(Color online) Propagation of bright one-soliton in 2-YO system for $k_1=1-i$, $c_1=1.5$, $c_2=1$, $\alpha_1^{(1)}=1$ and $\alpha_1^{(2)}=1.5$. Here $c_1,c_2>0$ and $\Theta <0$; hence the value $e^R$ becomes positive-definite only for $k_{1I} < 0$.}
\label{osfig1}
\end{figure}

\noindent{\textbf{Case (i): Positive nonlinearity coefficients} ($c_{\ell}>0$)}\\
\indent For the choice $c_{\ell}>0$, in order to get a nonsingular solution, we need $e^R$ to be positive-definite which requires $k_{1I}<0$ [see the expression for $e^R$ below Eq. (\ref{1sol})]. For better understanding, we have plotted the propagation of bright one-soliton arising in the 2-YO system in Fig. \ref{osfig1} for the choice $c_1,~c_2>0$. The quantities appearing in Fig. 1 and also in the rest of the figures are dimensionless. Here the solitons are localized in positive (negative) `$x$' axis at $t=-3$ ($t=3$).

\noindent{\textbf{Case (ii):Negative nonlinearity coefficients}($c_{\ell}<0$)}\\
\indent For this choice, $c_{\ell}<0$, the restriction on $e^R$ to be positive-definite for obtaining regular soliton requires $k_{1I}$ to be positive ($k_{1I}>0$). As the sign of $k_{1I}$ is reversed from that of the previous case, here the direction of propagation of the soliton is opposite to that of the previous case ($c_{\ell}>0$), for the same values of magnitude of $k_{1I}$. However, the intensity of the soliton will remain same in both cases for same values of $\alpha_1^{(\ell)}$, $k_{1R}$, and $|c_{\ell}|,~\ell=1,2,...,M$.\\

\noindent{\textbf{Case (iii):Mixed-type nonlinearity coefficients}\\
\indent When the nonlinearity coefficients $c_{\ell}$ admit mixed signs (say for instance, $c_{\ell} >0$ for $\ell=1,2,...,m,$ and $c_{\ell}<0$ for $\ell=m+1,m+2,...,M$), the formation of the nonsingular soliton solution requires $e^R>0$, which in turn depends on the signs of $k_{1I}$ and $-\sum_{\ell=1}^{M}(c_{\ell} |\alpha_1^{(\ell)}|^2) (\equiv \Theta$) (see below Eq. (\ref{1sol}b)). As a consequence of this, we can further divide this case into two subcases, namely, (a) $\Theta<0$ and $k_{1I}<0$ and (b) $\Theta>0$ and $k_{1I}>0$. It can be understood that $\Theta$ and $k_{1I}$ should possess the same sign to obtain nonsingular soliton solutions. Here, the $\Theta$ value will be different from that of the previous two cases (i) \& (ii), for the same choice of soliton parameters and for $c_{\ell}$ with same the magnitudes. Ultimately, the intensity of the soliton in this case will be different from the previous two cases, that are having same intensity, for a given set of soliton parameters with the magnitude of $c_{\ell}$ being the same. It is obvious that for different $c_{\ell}$ values the soliton amplitude will vary based on $|\Theta|$. However, the amplitude of the soliton appearing in the long-wave component remains the same in all three cases, as it is independent of $\alpha_1^{(\ell)}$ and $|c_{\ell}|$ parameters and is proportional to $k_{1R}^2$ only (see Eq. (\ref{1sol}b)).
\begin{widetext}\vspace{-0.51cm}
\begin{table}[b]\vspace{-0.5cm}
\caption{Bright one-soliton propagation in $M$-YO system}
\label{table1}
\begin{tabular}{|c|p{3.55cm}|p{2.6cm}|c|p{1.1cm}|p{4.15cm}|p{2.3cm}|}
  \hline
Case & ~~~~~~~~~Choice of $c_{\ell}$ & ~~Condition for & \multicolumn{3}{|c|}{SW soliton} & ~~~LW soliton  \\\cline{4-6}
     &                      & ~~regular soliton  & Amplitude  & Velocity  & ~~~~~~~~~~~Description  & \\ \hline
  (i) & $c_{\ell}>0$,   for $\ell=1,2,...,M$. & $k_{1I}<0$ & $2ik_{1R}{\alpha_1^{(\ell)}}\sqrt{|\frac{k_{1I}}{\Theta}|}$ & ~$2|k_{1I}|$ & Intensities are same but the & Intensities are\\\cline{1-5}
  (ii) & $c_{\ell}<0$,   for $\ell=1,2,...,M$. & $k_{1I}>0$ & $-2ik_{1R}{\alpha_1^{(\ell)}}\sqrt{|\frac{k_{1I}}{\Theta}|}$ & $-2|k_{1I}|$ & propagation directions are opposite in (i) \& (ii). & same in all three cases.\\\cline{1-6}
  (iii) & $c_{\ell}>0$,   for $\ell=1,2,...,m$, & (a) $k_{1I}<0$, $\Theta<0$ & $2ik_{1R}{\alpha_1^{(\ell)}}\sqrt{|\frac{k_{1I}}{\Theta}|}$ & ~$2|k_{1I}|$ & Intensities are different from & Velocity is equal\\\cline{3-5}
        & $c_{\ell}<0$,   for $\ell=m+1,m+2,...,M$. & (b) $k_{1I}>0$, $\Theta>0$ & $-2ik_{1R}{\alpha_1^{(\ell)}}\sqrt{|\frac{k_{1I}}{\Theta}|}$ & $-2|k_{1I}|$ & the cases (i) \& (ii). Velocity in sub-cases (a) \& (b) are similar to that of cases (i) \& (ii), respectively. &  to that of SW soliton appearing in the respective cases.\\
  \hline
\end{tabular}
\end{table}\end{widetext}

The propagation dynamics of the bright one-soliton discussed here is summarized in Table I. Note that, the value of $|\Theta|~(\equiv 2ik_{1R}{\alpha_1^{(\ell)}}\sqrt{|\frac{k_{1I}}{\Theta}|})$ for case (iii) is different from that of cases (i) and (ii) and so their intensities. In fact, the $|\Theta|$ value is different, in general, for the two subcases of (iii) too and can be made equal for particular $c_{\ell}$ values thereby rendering the possibility of same intensity solitons.

The above discussion on the one-soliton solution of the general $M$-YO system (\ref{model}) clearly shows that the choice of nonlinearity coefficients ($c_{\ell}$) predominantly determines the nature of soliton propagation, particularly in SW components, and influences the soliton amplitude and also the velocity indirectly.

\subsection{Bright, dark and anti-dark solitons}\label{secdark}
To illustrate the application of the present solution, we construct the asymptotic solution of system (\ref{eq1}), when the LSRI takes place, by inverting the soliton solutions (\ref{1sol}) using Eqs. (\ref{trans}) and (\ref{choice}) and expressing the resulting equations in terms of the original co-ordinates $Z$ and $T$ [by using the transformations given above Eq. (\ref{trans})]. One can also notice that the parameter $\varepsilon$ can be absorbed into the arbitrary soliton parameters $k_{1R}$ and $k_{1I}$, hence it plays no role in soliton dynamics. The study of these asymptotic solitons in 3-CNLS system (2) gives insight into the dynamics of interaction of three waves, when the LSRI condition is achieved. We find that the asymptotic solitons appearing in the $q_{\ell+1},~\ell=1,2$, components look similar to the soliton profiles of short waves $S^{(\ell)},~\ell=1,2$. On the other hand, the bright soliton appearing in the $L$ component will take different types of soliton profiles in its original coordinate (i.e., in the $q_1$ component): an anti-dark soliton which is nothing but a bright soliton pulse appearing in a constant background (i.e., bright soliton with nonzero amplitude at $T\rightarrow \pm \infty$), a dark soliton with single-well and double-well type structure, and a gray soliton (nonzero amplitude dip on a constant background), depending on the parameter $\sigma$ ($\equiv \sigma_{21}=\sigma_{31}$), which in turn depends on the nonlinearity coefficients $c_{\ell}$ as given by Eq. (\ref{choice}). Also, these asymptotic solitons propagate with common velocity in all three components. Particularly, we obtain anti-dark soliton (dark/gray soliton) in the $q_1$ component when the value of $\sigma$ is positive (negative). Such dark-bright, anti-dark--bright, and gray-bright solitons are possible for all three cases of $c_{\ell}$. Similar types of bright-dark and bright--anti-dark solitons have been reported in the 2-CNLS system with third-order dispersion effect \cite{frantzanti}. In our present study, we obtain such solitons in 3-CNLS equations even in the absence of third order dispersion. Here the existence of such anti-dark  solitons in system (\ref{eq1}) is a special feature of long-wave--short-wave resonance interaction. More interestingly, in a recent work \cite{3wave2} similar types of velocity-locked bright-bright-dark solitons are obtained in the context of resonant three-wave interaction systems in quadratic media and their special dynamics have been explored. Infact, the LSRI process is a special case of the three-wave interaction process \cite{threewave,3photonic,3wave2}. This suggests the possibility of such velocity-locked soliton like structures even in system (2), when such LSRI takes place, and also can be observed in cascaded nonlinear systems \cite{cascaded}.
\begin{figure}[h]
\centering\includegraphics[width=1.0\linewidth]{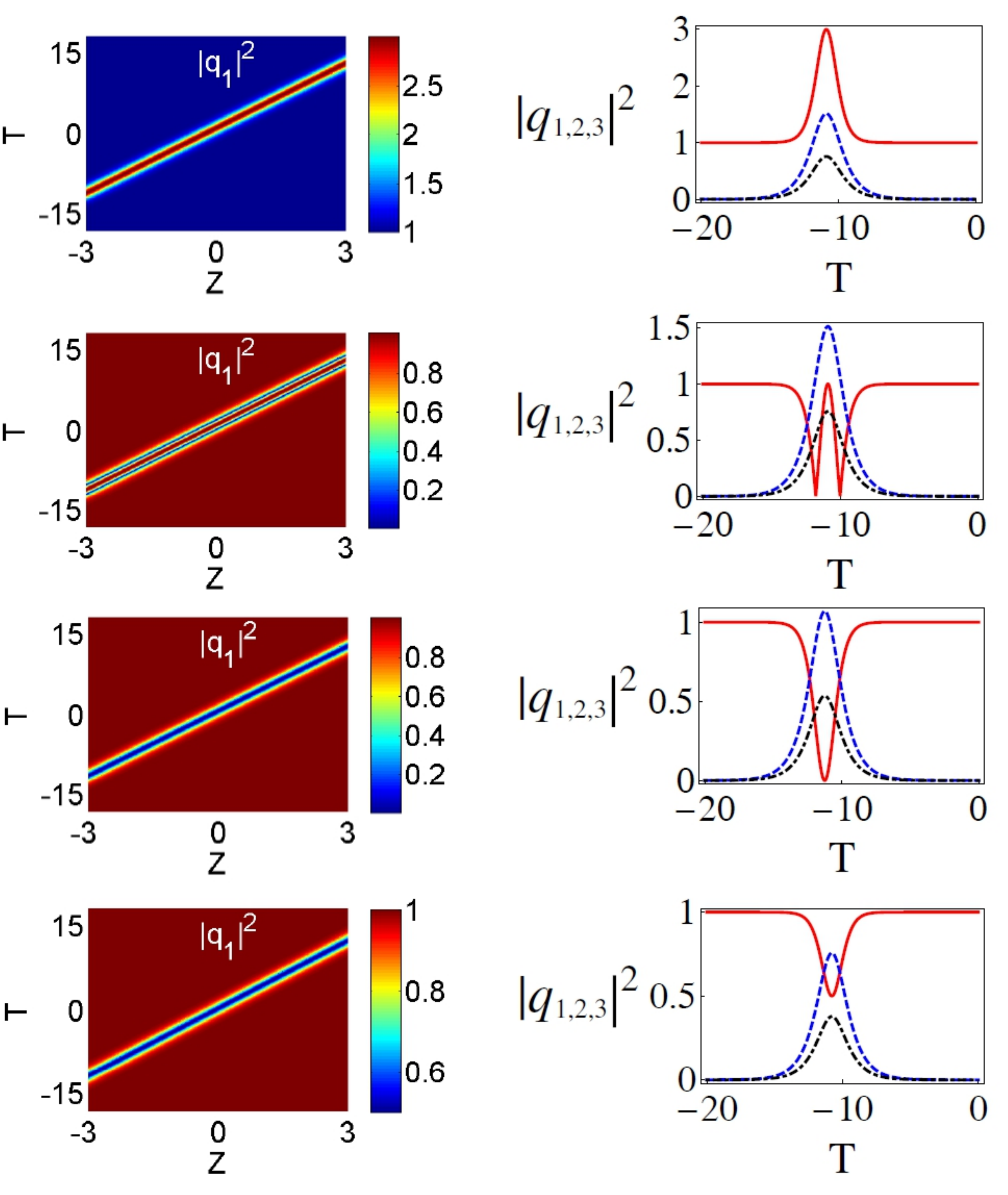}
\caption{(Color online) Different types of asymptotic solitons appearing in 3-CNLS system. From top to bottom: Anti-dark--bright, double-well-dark--bright, dark-bright, and gray-bright one-soliton. Right panels: $q_1$, solid line; $q_2$, dashed line; $q_3$, dot-dashed line.}
\label{bdosfig1}
\end{figure}

To elucidate the above points, we plot anti-dark--bright, dark-bright, and gray-bright solitons appearing in the 3-CNLS system (\ref{eq1}), when there occurs a resonant interaction, in Fig. \ref{bdosfig1}. In the left panels of Fig. \ref{bdosfig1}, from top to bottom, we have shown the color density plots corresponding to the $q_1$ component, which supports the (a) anti-dark--bright soliton, (b) dark-bright soliton with dark component admitting double-well type structure, (c) dark-bright soliton and (d) gray-bright soliton for various choices of $\sigma$ parameter. Also, their typical two-dimensional intensity plots along with that of $q_2$ and $q_3$ components at $Z=-3$ are depicted on the right panels for better illustration. The parameters in the top panels of Fig. \ref{bdosfig1} are chosen as $k_1=1-i$, $\alpha_1^{(1)}=1$, $\alpha_1^{(2)}=0.5$, $\sigma=1$, $\sigma_{11}=2$, $\sigma_{12}=-1.5$ and $\sigma_{13}=-1$. For the remaining panels of Fig.  \ref{bdosfig1}, the same values have been assigned for $\sigma_{11}, \sigma_{12}$ and $ \sigma_{13}$ (respectively, $2, 1.5 $ and $1$), but the numerical values for $\sigma$  in the second, third, and bottom panels are respectively $-1, -2$ and $-4$, with $k_1$,  $\alpha_1^{(1)}$, and $\alpha_1^{(2)}$ being the same as that of the top panels. Especially, the $\sigma$ parameter defines the nature of asymptotic soliton profile in the $q_1$ component. These choices and their corresponding profiles are given in Table II.
\begin{table}[h]
\label{table2}
\begin{center}
\caption{Profiles of asymptotic solitons.} 
\begin{tabular}{@{}|c|c|}
\hline
Choices of $\Pi~\left(=\frac{2k_{1R}^2}{\sigma}\right)$ & Nature of soliton in $q_1$ \\
\hline
$\Pi > 0$ & anti-dark soliton\\\hline
$0>\Pi >-1$ & gray soliton \\\hline
$ \Pi \leq -1$ & dark soliton \\
\hline
\end{tabular}
\end{center}
\end{table}

\subsection{Bright two-soliton solution}
The explicit expression for bright two-soliton solution of $M$-LSRI system (\ref{model}) can be obtained from Eq. (\ref{nsol}) for $N=2$ and can be cast as
\bes\bea
S^{(\ell)}&=&\frac{1}{f}\left(\alpha_1^{(\ell)} e^{\eta_1}+\alpha_2^{(\ell)} e^{\eta_2}+e^{\eta_1+\eta_1^{*}+\eta_2+\delta_{1}^{(\ell)}}\right.\nonumber\\ &&\left. ~~~+e^{\eta_2+\eta_2^{*}+\eta_1+\delta_{2}^{(\ell)}}\right),~ \ell=1,2,3,...,M,~~\\
L&=&2\frac{\partial^2}{\partial x^2}(\ln{f}),
\eea
where
\bea
f&=&1+e^{\eta_1+\eta_1^{*}+R_1}+e^{\eta_1+\eta_2^{*}+\delta_0}+e^{\eta_2+\eta_1^{*}+\delta_0^{*}}\nonumber\\ &&~~+e^{\eta_2+\eta_2^{*}+R_2}+e^{\eta_1+\eta_1^{*}+\eta_2+\eta_2^{*}+R_3}.
\eea
Various other quantities appearing in the above solution (\ref{2sol}) are defined below:
\bea
\eta_j&=&k_j x+ik_j^2 t,  \quad j=1,2,~~~~~\\
e^{R_j}&=&\frac{\kappa_{jj}}{k_j+k_j^*},~ e^{\delta_0}=\frac{\kappa_{12}}{k_1+k_2^*},~ e^{\delta_0^*}=\frac{\kappa_{21}}{k_2+k_1^*},~~~~ \\
e^{\delta_{1}^{(\ell)}}&=&\frac{(k_1-k_2)(\alpha_1^{(\ell)}\kappa_{21}-\alpha_2^{(\ell)}\kappa_{11})}{(k_1+k_1^*)(k_2+k_1^*)},~~ \ell=1,2,~~~~\\
e^{\delta_{2}^{(\ell)}}&=&\frac{(k_1-k_2)(\alpha_1^{(\ell)}\kappa_{22}-\alpha_2^{(\ell)}\kappa_{12})}{(k_2+k_2^*)(k_1+k_2^*)},~~ \ell=1,2,~~~~\\
e^{R_3}&=&\frac{|k_1-k_2|^2(\kappa_{11}\kappa_{22}-\kappa_{12}\kappa_{21})}{(k_1+k_1^*)(k_2+k_2^*)|k_1+k_2^*|^2},\\
\kappa_{pj}&=&\frac{\ds\sum_{\ell=1}^M\left(c_{\ell}\alpha_p^{(\ell)}\alpha_j^{(\ell)*}\right)}{2i(k_p^2-k_j^{*2})}, \quad p,j=1,2.\label{kappa}
\eea\label{2sol}\ees

The above bright two-soliton solution of the $M$-component YO system (\ref{model}) is characterized by ($2M+2$) number of complex parameters ($\alpha_j^{(\ell)}$ and $k_j$, $j=1,2;~\ell=1,2,3,...,M$). The nature of the above two-soliton solution (singular or nonsingular) explicitly depends on the value of `$f$', which results in nonsingular solutions for $f > 0$; otherwise (for $f<0$) the solution becomes singular. Particularly, the nature of the solution is based on the values of $\kappa_{pj}$, $p,~j=1,2$: the nonsingular solutions result for $\kappa_{11},~\kappa_{22}>0$ and $\kappa_{11}\kappa_{22}-\kappa_{12}\kappa_{21}>0$, for which $f$ becomes positive-definite. In the present system, the arbitrariness of nonlinearity coefficients ($c_{\ell}$) and the polarization parameters predominantly determine the nature of soliton collision and allow the $M$-YO  system to host rich interaction dynamics of solitons as will be demonstrated in the following section.

\section{Soliton collisions}\label{collisions}
Generally, multicomponent soliton collisions display several distinct features \cite{rkpre,tkprl,tkpre,tkjpa,tklsri,tklsribd} due to the additional freedom resulting from the multicomponent nature. The detailed dynamics of soliton collisions described by two-soliton solution (\ref{2sol}) can be explored by analyzing the solution \cite{tkpre,tkjpa,tklsri,tklsribd} in asymptotic limits. In this section, we discuss the collision among two bright solitons of the 2-component YO system comprising two short waves and a long wave. This study can be directly extended to the $M$-YO  system with more than two short waves. The dependence of the existence of the nonsingular solution on the velocity $k_{jI}$, where $j=1,2$, restricts the two colliding solitons to propagate in the same direction. In fact, solitons traveling opposite to each other is not at all possible in the present system (\ref{model}) as this choice will result in singular solution. Thus, system (\ref{model}) can only feature overtaking collision and one can not have head-on collision in system (\ref{model}). However such head-on collision is possible in system (\ref{eq1}), even when LSRI takes place, as will be shown latter in this section. For performing the asymptotic analysis, we consider the choice $k_{1R},~k_{2R}>0$ and $k_{1I},~k_{2I}<0$ (or $k_{1I},~k_{2I}>0$) without loss of generality. To be short in the presentation, we do not give the complete expressions for asymptotic analysis since the procedure has been explained in detail in several earlier works \cite{tkpre,tkjpa,tklsri,tklsribd}. Here, the nature of nonlinearities (i.e., mainly the sign of nonlinearity coefficients $c_{\ell}$) plays an important role in defining the direction of propagation of solitons. As in the one-soliton case, here also, we discuss the soliton collisions for the three choices of $c_{\ell}$.\\

\noindent{\bf Case (i): Positive nonlinearity coefficients ($c_{\ell}>0$)}\\
The solitons of one-dimensional $M$-YO  system (\ref{model}) undergo novel energy-sharing collisions, like collision of bright solitons in multicomponent Manakov (CNLS) systems, and such collision scenario is shown in Fig. \ref{sc1fig}. In Fig. \ref{sc1fig}, one can observe that the intensity of soliton $s_1$ ($s_2$) is suppressed (enhanced) in the $S^{(1)}$ component and it gets enhanced (suppressed) in the $S^{(2)}$ component after collision with soliton $s_2$ ($s_1$). To elucidate the understanding of such collision, we perform an asymptotic analysis for this choice of nonlinearity coefficient and obtain the following equation relating the amplitudes of solitons after and before collision in the short-wave components:
\bes\bea
A_j^{(\ell)+} = T_j^{(\ell)} A_j^{(\ell)-}, \qquad j=1,2, \qquad \ell=1,2,
\eea
where $T_j^{(\ell)}$'s are transition amplitudes
\bea
T_1^{(\ell)} = \frac{1-\lambda_1}{\sqrt{1-\lambda_1 \lambda_2}}\left(\frac{(k_1-k_2)(k_1^*+k_2)}{(k_1^*-k_2^*)(k_1+k_2^*)} \right)^{1/2},\\
T_2^{(\ell)} = \frac{\sqrt{1-\lambda_1 \lambda_2}}{1-\lambda_2}\left(\frac{(k_1^*+k_2)(k_1^*-k_2^*)}{(k_1+k_2^*)(k_1-k_2)} \right)^{1/2}.
\eea\label{asymp}\ees
Here $\lambda_1=\frac{\alpha_2^{(\ell)} \kappa_{12}}{\alpha_1^{(\ell)} \kappa_{22}}$ and $\lambda_2=\frac{\alpha_1^{(\ell)} \kappa_{21}}{\alpha_2^{(\ell)} \kappa_{11}}$, where $\kappa_{pj},~p,j=1,2$ are defined in (\ref{kappa}). From the above equations it follows that, in general, for arbitrary $\alpha_j^{(\ell)}$ values, the transition amplitudes will not be unimodular and this results in the energy-sharing collision of solitons in the SW components. In this type of collisions, with positive nonlinearity coefficients ($c_{\ell}>0$), the energy in individual components as well as the total energy among all the components is conserved. We refer to this collision process as a type-I energy-sharing collision. Note that, the opposite kind of switching to that of Fig. \ref{sc1fig}, i.e., enhancement (suppression) of intensity for $s_1$ ($s_2$) in the $S^{(1)}$ component with commensurate changes in $S^{(2)}$ component, can also occur. Thus for the 2-YO system the type-I energy-sharing collision can take place in two different ways. However, the solitons in the LW component undergo only elastic collision, as their amplitudes do not depend on $\alpha_j^{(\ell)}$ parameters. Additionally, a given soliton $s_j$, $j=1,2$, experiences a phase-shift (say, $\Phi_j$, $j=1,2$) after collision, which is the same in all the components (both LW and SW components) and  is given by
\bea
\Phi_1&=&-\Phi_2=\ln\left(\left|\frac{k_1-k_2}{k_1+k_2^*}\right| \sqrt{1-\lambda_1 \lambda_2} \right).
\label{phase}\eea
These phase-shifts ultimately result in a change in the relative separation distance between the colliding solitons. The relative separation distances between the solitons ($s_1$ and $s_2$) before and after collision are found to be $t_{12}^{-}=\frac{(R_3-R_1)k_{1R}-R_1 k_{2R}}{2k_{1R}k_{2R}}$ and $t_{12}^{+}=\frac{R_2 k_{1R}-(R_3-R_2)k_{2R}}{2k_{1R}k_{2R}}$, respectively. Thus the change in relative separation distance $\Delta_{12}=t_{12}^{-}-t_{12}^{+}=\frac{k_{1R}+k_{2R}}{k_{1R}k_{2R}}\Phi_1$.
\begin{figure}[h]
\centering\includegraphics[width=1.0\linewidth]{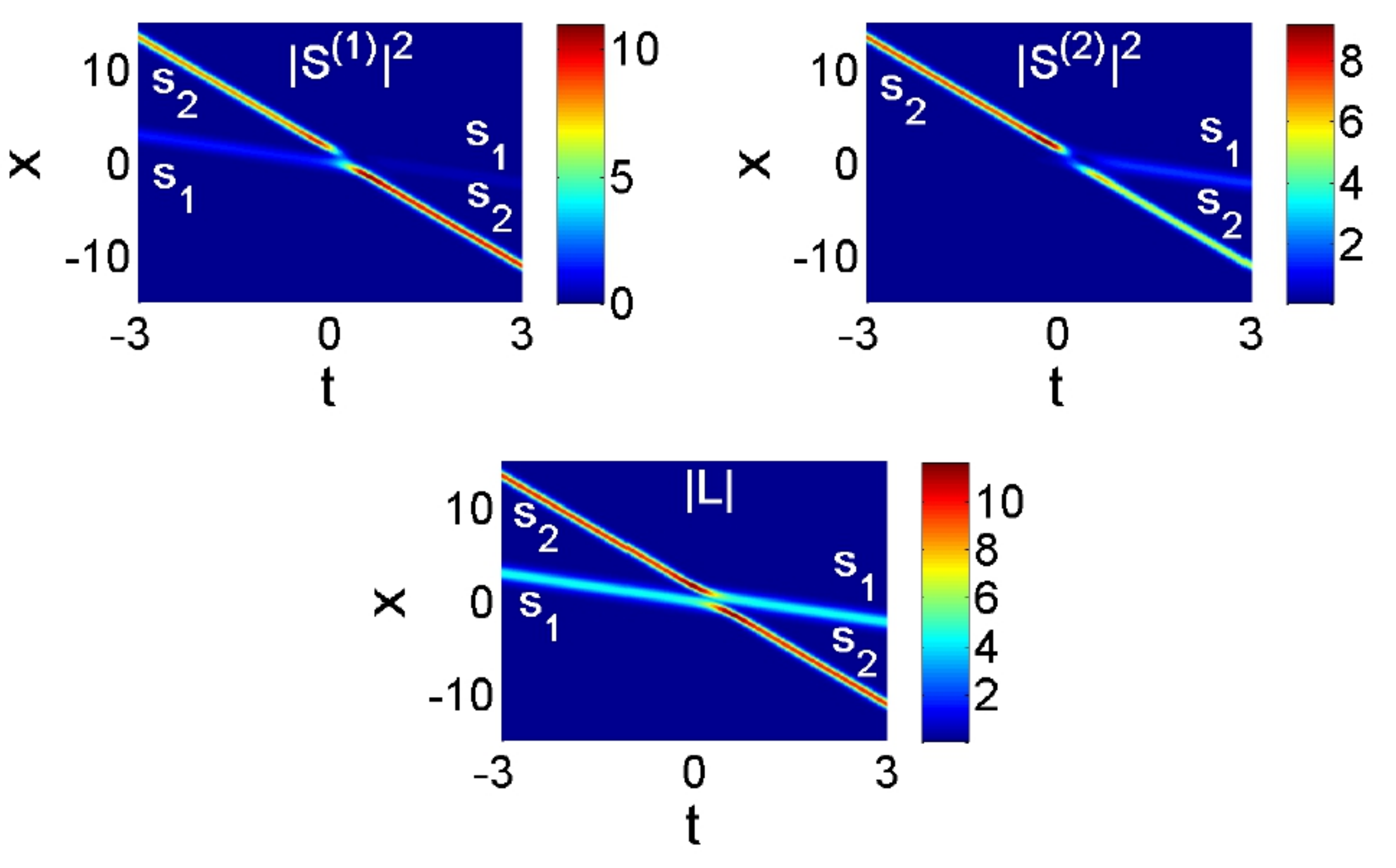}
\caption{(Color online) Type-I energy sharing collision in 2-YO system for $c_1,c_2>0$. Here $c_1=3$, $c_2=2$, $k_1=1.5-0.5i$, $k_2=2.2-2i$, $\alpha_1^{(1)}=2.4$, $\alpha_2^{(1)}=0.8$, $\alpha_1^{(2)}=0.7$, and $\alpha_2^{(2)}=0.6$.}
\label{sc1fig}
\end{figure}

The elastic collision of solitons results for the choice $\frac{\alpha_1^{(1)}}{\alpha_2^{(1)}} = \frac{\alpha_1^{(2)}}{\alpha_2^{(2)}}$, for which the transition amplitudes become unimodular, i.e., $|T_j^{(\ell)}|^2=1$. This choice is similar to the multicomponent Manakov system \cite{rkpre,tkprl,tkpre}. Here, the amplitudes of both solitons ($s_1$ and $s_2$) remain unaltered after collision in all three components (2-SW and 1-LW components). But they suffer a phase-shift after collision as given by (\ref{phase}). Such an elastic collision is shown in Fig. \ref{elasfig1}. For this choice, $c_{\ell}>0$, the solitons are localized in positive $x$ axis before collision ($t=-3$) and in negative $x$ axis after collision ($t=3$).
\begin{figure}[h]
\centering\includegraphics[width=1.0\linewidth]{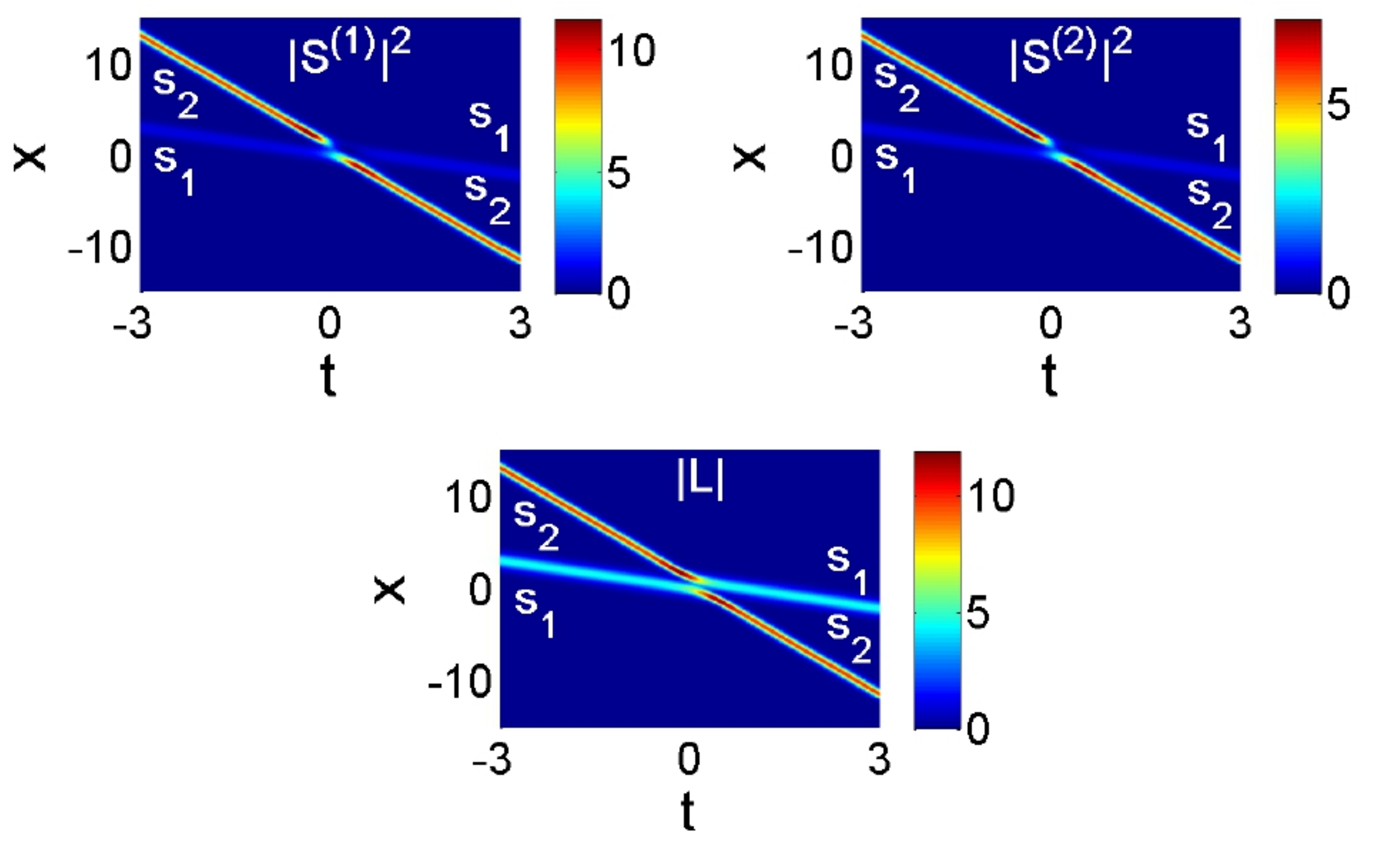}
\caption{(Color online) Elastic collision of two bright solitons in 2-YO system for $c_1,c_2>0$. The parameters are $c_1=3$, $c_2=2$, $k_1=1.5-0.5i$, $k_2=2.2-2i$, $\alpha_1^{(1)}=\alpha_2^{(1)}=2$, and $\alpha_1^{(2)}=\alpha_2^{(2)}=1.6$.}
\label{elasfig1}
\end{figure}

Next, it is of interest to discuss this type-I energy-sharing collision scenario in system (\ref{eq1}), when LSRI takes place, by inverting the two-soliton solution with the aid of expressions (\ref{trans}) and (\ref{choice}). Here also the solitons appearing in the $q_j,~j=2,3,$ components exhibit both elastic collision and energy-sharing collision of type-I, while the solitons in the $q_1$ component undergo elastic collision only. We have shown in Sec. \ref{secdark} that when LSRI occurs the LW solitons of system (2) can admit rich profile structures. It would be interesting to analyze the collision of such solitons for $c_{\ell}>0$. We notice that there are no dramatic profile changes in the $q_2$ and $q_3$ components; rather there appear only bright solitons and they undergo both elastic collision and type-I energy-sharing depending on the polarization parameters. Typical type-I energy-sharing collision in the $q_2$ and $q_3$ components is shown in the bottom panels of Fig. \ref{dbscfig1}. However, in the case of the $q_1$ component the asymptotic soliton profiles take interesting forms but the elastic nature of the collision still prevails. We have demonstrated such collision dynamics in the top panels of Fig. \ref{dbscfig1}. It is interesting to note that the $q_1$ component supports novel types of solitons arising due to the coexistence of both bright and dark parts in the same soliton with nonzero background (top panels of Fig. \ref{dbscfig1}). One can also obtain the soliton collisions for various other forms of solitons in $q_1$ resulting for different choices of $\sigma$ as shown in Fig. \ref{dbscfig2}. In spite of admitting such diverse soliton profiles in the $q_1$ component the solitons undergo only elastic collision. Another salient feature of soliton collision in system (2), when LSRI takes place is that the solitons can undergo both head-on and overtaking collisions. We believe that this type of novel soliton dynamics in 3-CNLS would give new directions in exploring their applications. One can also extend this study to a general $M$-YO system (1).
\begin{figure}[h]
\centering\includegraphics[width=1.0\linewidth]{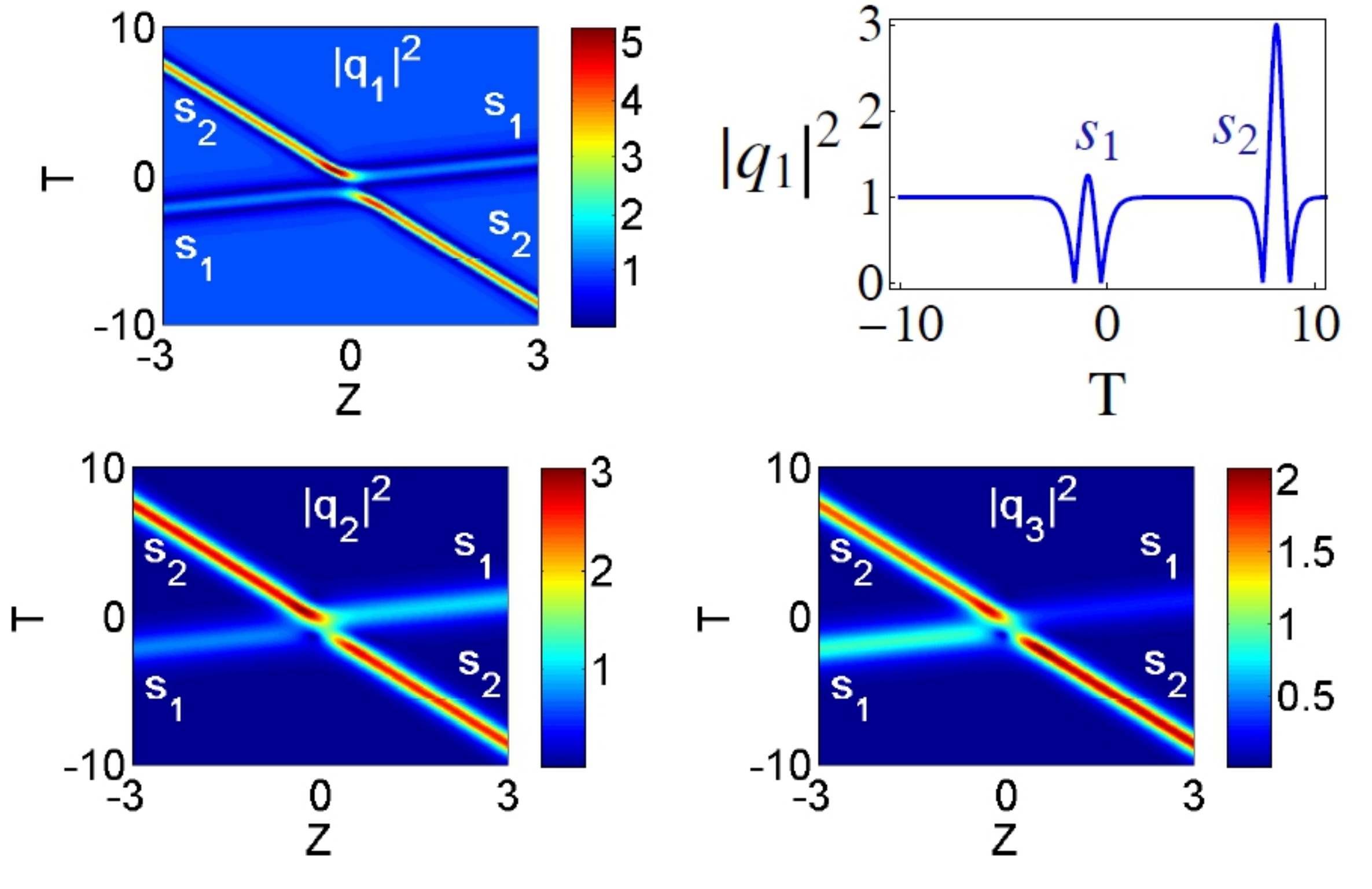}
\caption{(Color online) Elastic collision of novel bright-dark asymptotic solitons in $q_1$ component (top panels) and type-I energy-sharing collision of asymptotic solitons in $q_2$ and $q_3$ components (bottom panels) of 3-CNLS system (2). Here the soliton parameters are $k_1=1.5-0.5i$, $k_2=2-2i$, $\alpha_1^{(1)}=2.4$, $\alpha_2^{(1)}=0.8$, $\alpha_1^{(2)}=0.7$, $\alpha_2^{(2)}=0.6$ and the system parameters are $\sigma=-2$, $\sigma_{11}=2$, $\sigma_{12}=1.5$, $\sigma_{13}=1$.}
\label{dbscfig1}
\end{figure}

\begin{figure}[h]
\centering\includegraphics[width=1.0\linewidth]{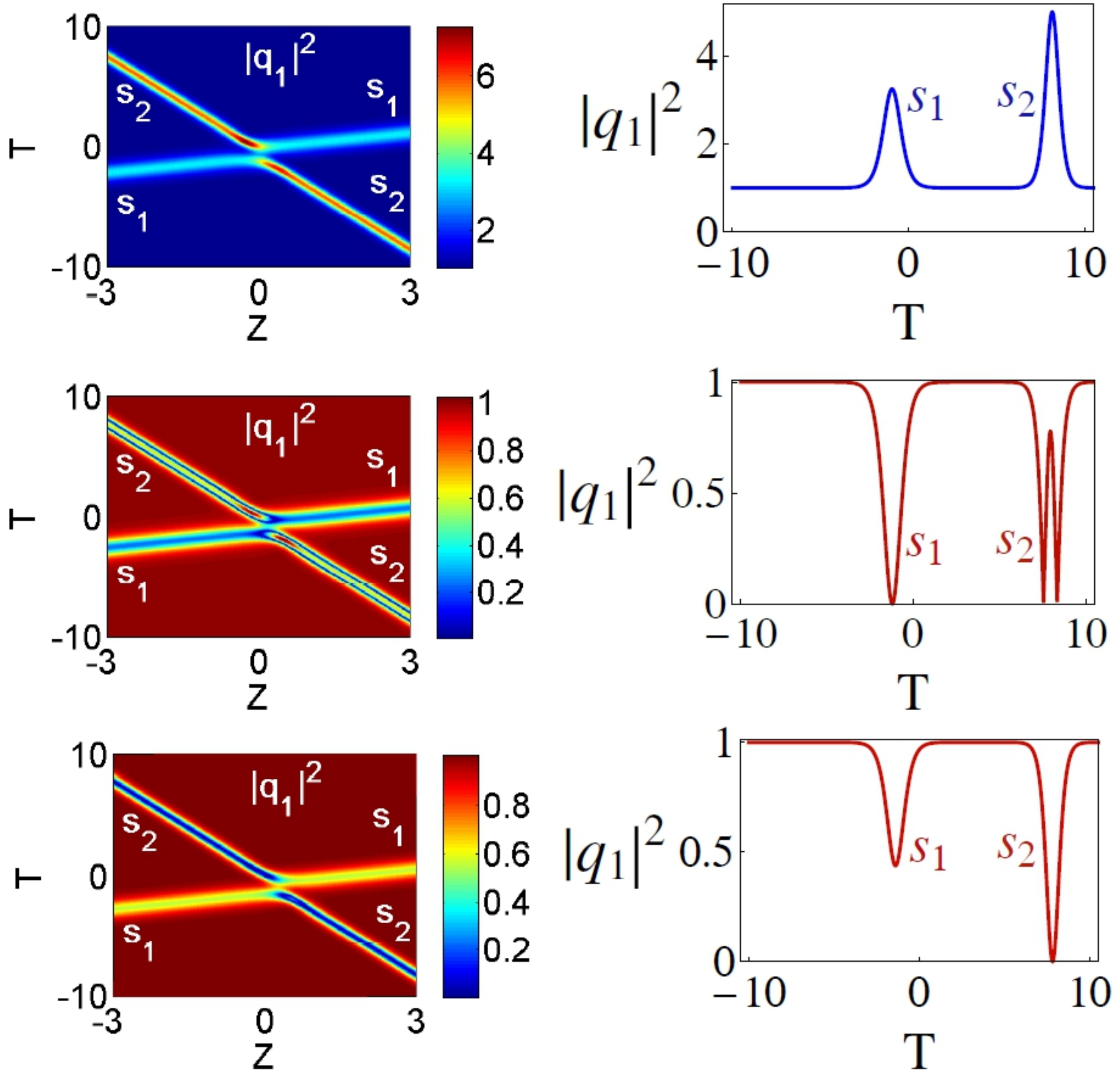}
\caption{(Color online) Elastic collision of asymptotic solitons in $q_1$ component of 3-CNLS system (2). From top to bottom: anti-dark soliton collision; collision between single- and double-well type dark solitons; collision between a dark and a gray soliton. Right panels: Soliton profile at $Z=-3$. Soliton parameters are same as in Fig. \ref{dbscfig1} and the system parameters are chosen as (top panels) $\sigma=1$, $\sigma_{11}=2$, $\sigma_{12}=-1.5$, $\sigma_{13}=-1$; (middle panels) $\sigma=-4$, $\sigma_{11}=2$, $\sigma_{12}=1.5$, $\sigma_{13}=1$; and (bottom panels) $\sigma=-8$, $\sigma_{11}=2$, $\sigma_{12}=1.5$, $\sigma_{13}=1$.}
\label{dbscfig2}
\end{figure}
~\\
\noindent{\bf Case (ii):Negative nonlinearity coefficients($c_{\ell}<0$)}\\
The soliton collision scenario corresponding to the $M$-YO  system with negative nonlinearity coefficients ($c_{\ell}<0$) is similar to that of $c_{\ell}>0$ case. The amplitudes of colliding solitons are exactly the same as the previous case, for the same choice of parameters except for $c_{\ell}$, which is now $c_{\ell}<0$, but there is a change in the direction of propagation compared to that of Figs. \ref{sc1fig} and \ref{elasfig1}. Also, in the original coordinate, the soliton collision in the 3-CNLS system looks similar to case (i) discussed above.\\

\noindent{\bf Case (iii): Mixed-type nonlinearity coefficients}\\
For mixed signs of $c_{\ell}$, the solitons in the SW component display a dramatic change in the collision behavior. This different type of energy-sharing collision is shown in Fig. \ref{sc2fig}. In Fig. \ref{sc2fig}, the amplitude of soliton $s_1$ ($s_2$) is enhanced (suppressed) in both short-wave components, while the solitons in the long-wave component reappears elastically after collision. Here the energy (or amplitude) of both solitons in LW component is the same after collision with a phase-shift. Here also, the amplitudes of solitons before and after collision are related by the same expressions (\ref{asymp}) and the phase-shifts are given by (\ref{phase}), but now in the expression $\kappa_{pj}$ [see Eq. (\ref{kappa})], $c_{\ell}$ take mixed signs rather than all $c_{\ell}$ being positive. In such collision, the energy in individual mode is conserved and the difference in energy, not total energy, between SW components is also conserved. As a consequence of this the nature of switching of intensities (energy) for a given soliton in both short-wave components is the same, while in a given component the colliding solitons $s_1$ and $s_2$ experience the opposite kind of energy switching. That is, the amplitude of soliton $s_1$ is enhanced (or suppressed) while the amplitude of soliton $s_2$ gets suppressed (or enhanced) after collision in both short-wave components. Note that the reverse collision scenario to that of Fig. \ref{sc2fig}, i.e., suppression (enhancement) of intensity for $s_1$ ($s_2$) is also possible. We refer to such collision resulting in the same kind of switching behavior for a particular soliton in both SW components as a type-II energy-sharing collision. This type of energy-sharing collision is a special feature of the $M$-YO  system with mixed signs of nonlinearity coefficients. Such collision has been observed earlier in the mixed CNLS system \cite{tkpre}. This collision scenario can be profitably used for the amplification of a particular soliton in both short-wave components by treating the other colliding soliton as a probe soliton, that will enable such switching.
\begin{figure}[h]
\centering\includegraphics[width=1.0\linewidth]{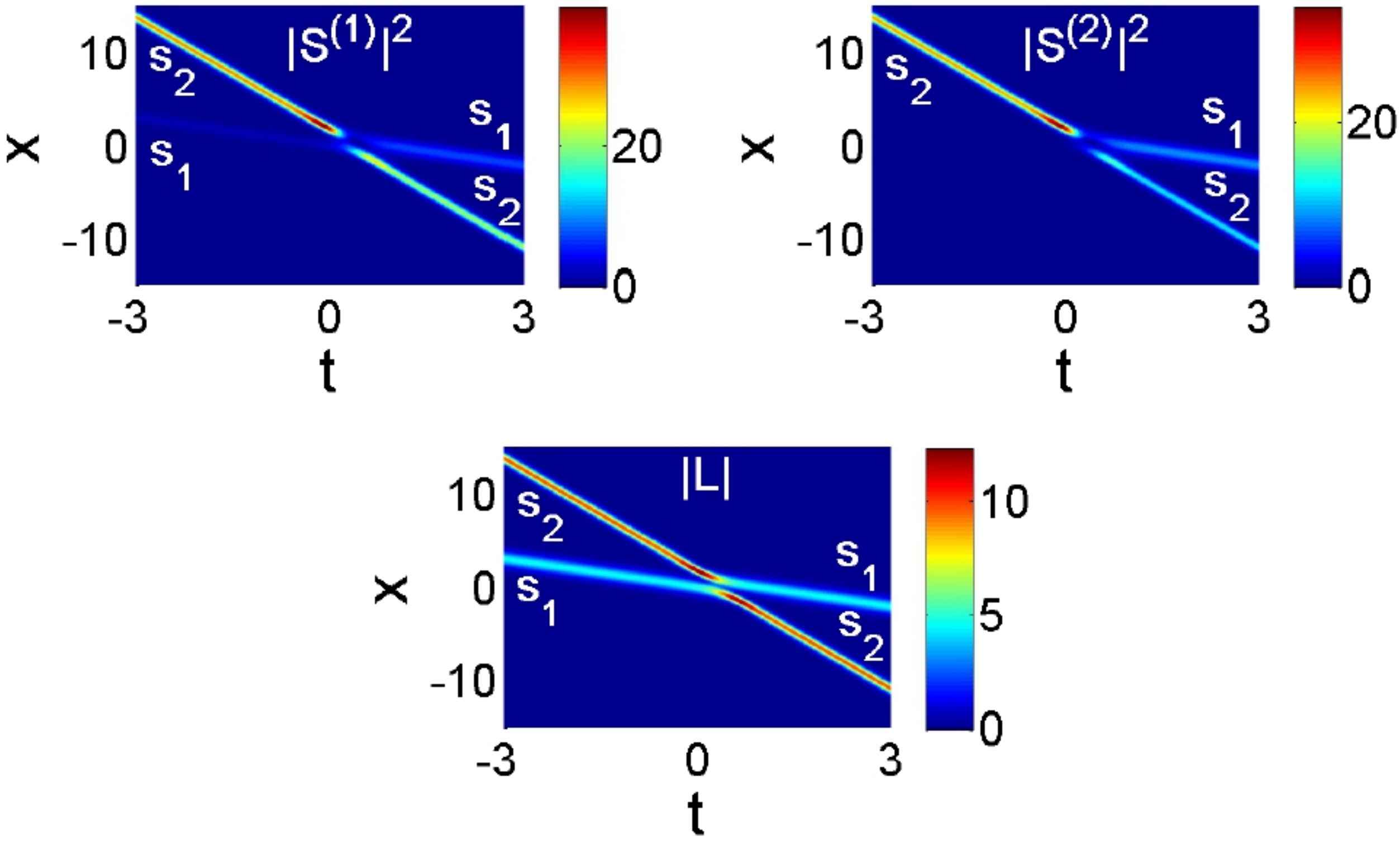}
\caption{(Color online) Type-II energy-sharing collision in 2-YO system for $c_1=3$, $c_2=-2$, $k_1=1.5-0.5i$, $k_2=2.2-2i$, $\alpha_1^{(1)}=2.4$, $\alpha_2^{(1)}=0.8$, $\alpha_1^{(2)}=0.7$, $\alpha_2^{(2)}=0.6$.}
\label{sc2fig}
\end{figure}

In addition to the type-II energy-sharing collision, the solitons also undergo elastic collision without any change in their intensities in the SW components, for particular choice of polarization parameters $\alpha_j^{(\ell)}$, as well as in the LW component and they exhibit only a phase-shift.

As in case (i), here also one can study the nature of asymptotic soliton collision in system (\ref{eq1}) with the aid of present analysis on 2-YO system in the presence of mixed nonlinearities. The results are similar to case (i), except for the important difference that now the energy-sharing collision in $q_2$ and $q_3$ components is of type II.

\subsection{Three-soliton collision}\label{3sol}
It is of further interest to investigate the collision dynamics of multiple bright solitons, involving more than two solitons in $M$-YO  system (\ref{model}). In this subsection, we briefly point out the important features of the three-soliton collision of the two-component YO system. The exact three-soliton solution can be obtained from Eq. (\ref{nsol}) by substituting $N=3$ and $M=2$. One can perform the asymptotic analysis for a detailed study on three soliton collisions. Through such analysis we identify that the three-soliton collision is pairwise and no multiparticle effect takes place. Also, the solitons undergo energy-sharing collisions of type I and type II depending on the choice of $c_{\ell}$ parameters and elastic collisions depending on the $\alpha_j^{(\ell)}$ parameters. These type-I and type-II energy sharing collision of three bright solitons in the 2-YO system are shown in Fig. \ref{3stype12}. The elastic collision of three bright solitons occurs for the specific choice of polarization parameters satisfying the relation $\alpha_1^{(1)}:\alpha_2^{(1)}:\alpha_3^{(1)}=\alpha_1^{(2)}:\alpha_2^{(2)}:\alpha_3^{(2)}$ accompanied by a phase-shift. On the other hand, the energy sharing collisions of both types take place for all other choices of the $\alpha_j^{(\ell)}$ parameters, where $j=1,2,3$ and $\ell=1,2$.
\begin{figure}[h]
\centering\includegraphics[width=1.0\linewidth]{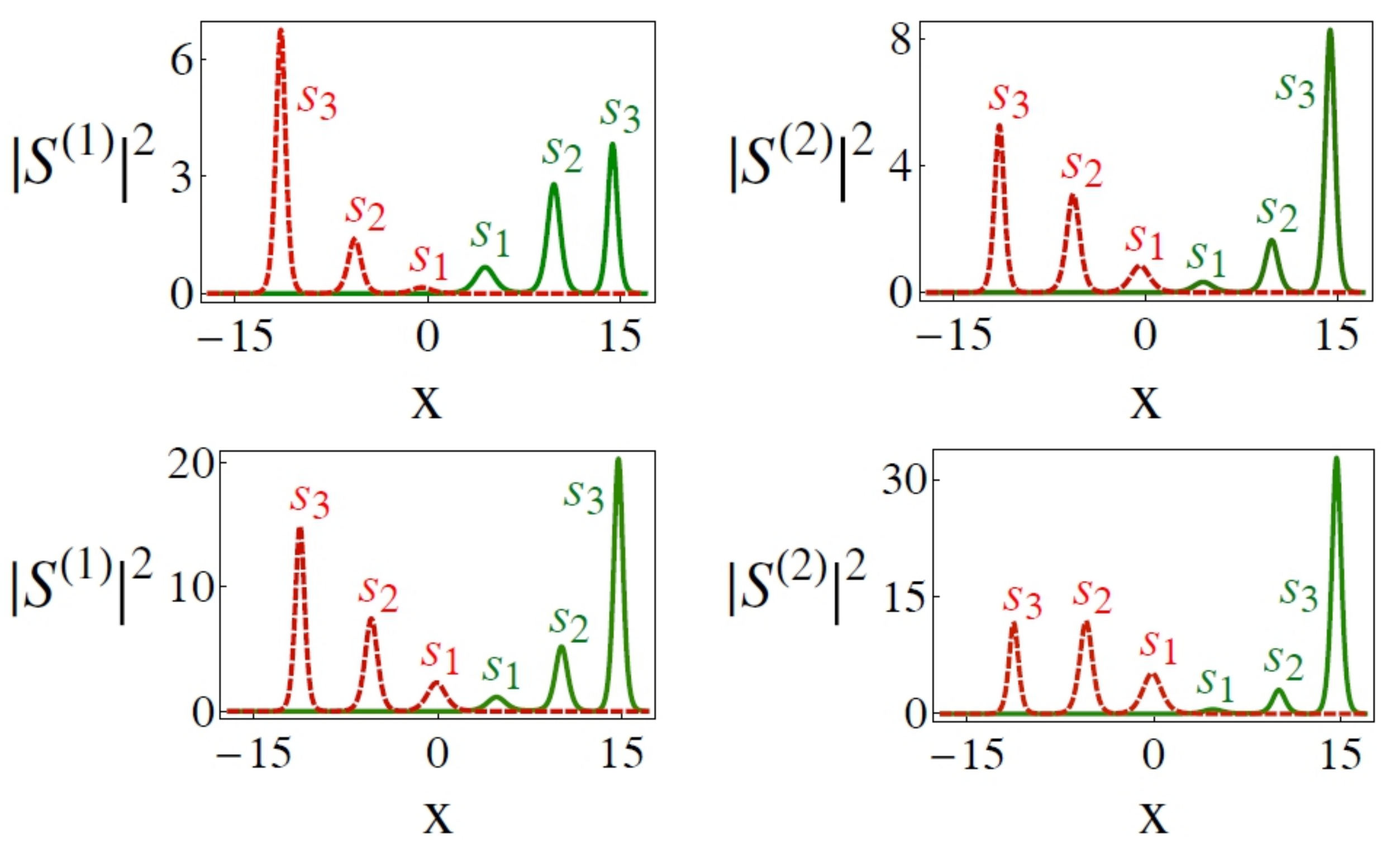}
\caption{(Color online) Energy-sharing collision of three bright solitons of type-I (top panels) and type-II (bottom panels) in the short-wave components of 2-YO system. The solid and dashed lines represent the solitons before ($t=-4$) and after ($t=4$) collision, respectively. The soliton parameters are $k_1=1-0.5i$, $k_2=1.5-i$, $k_3=2-1.5i$, $\alpha_1^{(1)}=1$, $\alpha_2^{(1)}=1.5$, $\alpha_3^{(1)}=1.71$, $\alpha_1^{(2)}=0.7$, $\alpha_2^{(2)}=1.1$, and $\alpha_3^{(2)}=1.51$, while the system parameters are $c_1=2$ and $c_2=2$ for top panels, and $c_1=2$ and $c_2=-0.5$ for bottom panels.}
\label{3stype12}
\end{figure}

It is interesting to note that in the type-I energy-sharing collision process of three solitons $\alpha_j^{(\ell)}$ parameters can be chosen in such a way that the state of any one of the solitons can be restored after two consecutive collisions with two solitons in the SW components. This type of state restoration is a special signature of multisoliton collision in Manakov-type soliton collisions \cite{tkpre} which we observe now in the $M$-YO  system (\ref{model}). This state restoration of soliton $s_1$ after two consecutive collisions with solitons $s_2$ and $s_3$, respectively, is demonstrated in Fig. \ref{restore}. The solitons $s_2$ and $s_3$ experience switching in their intensity after collision whereas $s_1$ remains unaltered. Here all three solitons experience a phase-shift. This will ordain the present multicomponent Yajima-Oikawa system as a suitable candidate for performing inverse operations in the context of soliton collision based optical computing.
\begin{figure}[h]
\centering\includegraphics[width=1.0\linewidth]{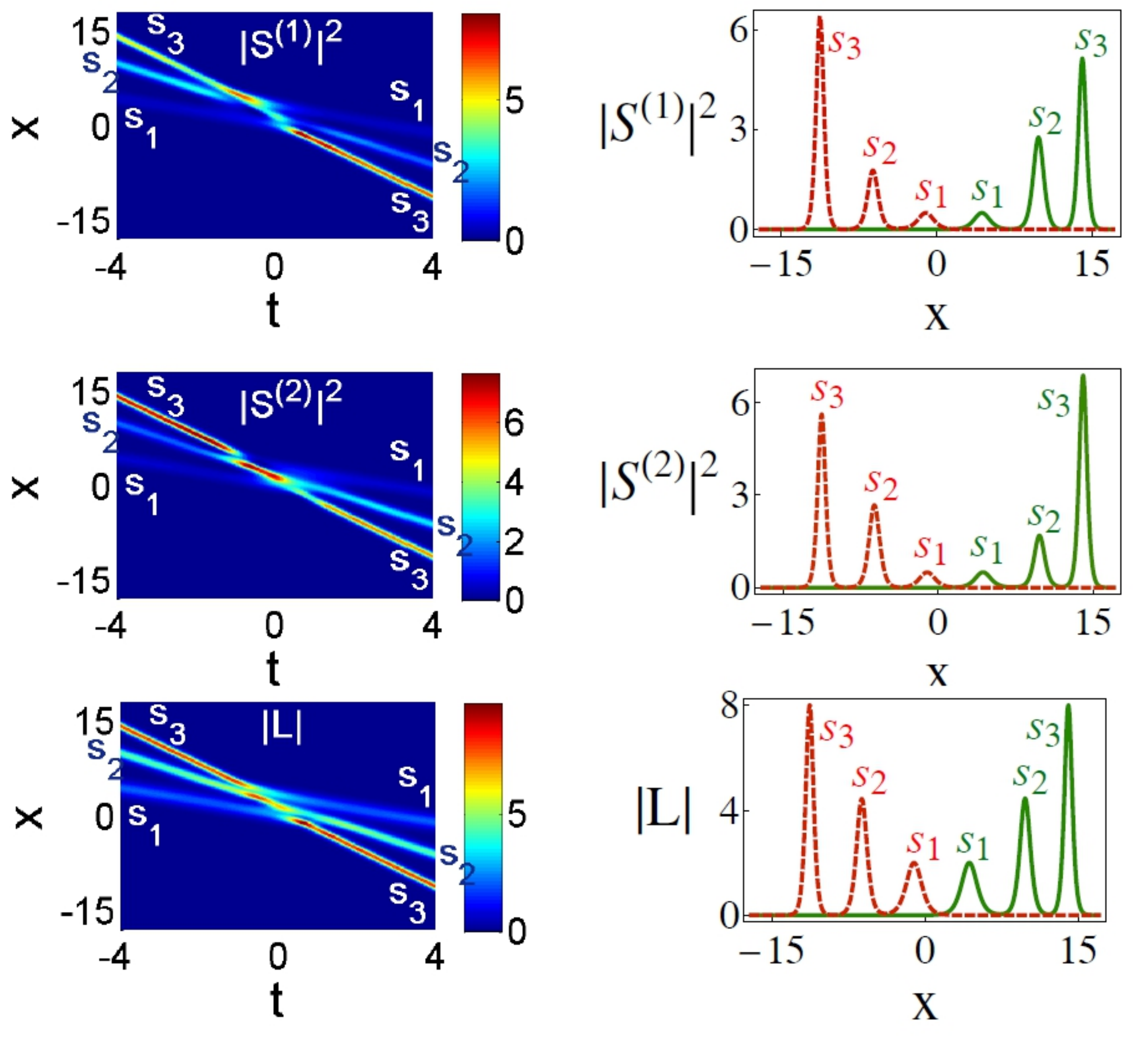}
\caption{(Color online) State restoration of soliton $s_1$ in the short-wave components of 2-YO system. The solid and dashed lines represent the solitons before ($t=-4$) and after ($t=4$) collision, respectively. The parameters are $c_1=2$, $c_2=2$, $k_1=1-0.5i$, $k_2=1.5-i$, $k_3=2-1.5i$, $\alpha_1^{(1)}=\alpha_2^{(1)}=\alpha_1^{(2)}=1$, $\alpha_2^{(2)}=0.5+i$, $\alpha_3^{(1)}=1.19$, and $\alpha_3^{(2)}=0.5+i$.}
\label{restore}
\end{figure}

\section{Conclusion}\label{conclusion}
To conclude, we have considered the multicomponent one-dimensional Yajima-Oikawa system describing the resonant interaction of multiple short-waves with a long wave. By using the asymptotic reduction method, we derive a two-component YO system from the 3-CNLS system (2) to illustrate the physical significance of the multicomponent YO system (1). The study on the integrability nature of the $M$-YO  system is carried out by performing the Painlev\'e analysis and the system is found to be integrable for arbitrary nonlinearity coefficients $c_{\ell}$. More interestingly, different sign choices of $c_{\ell}$ result in solitons with different nature and also in a dramatic difference in their collision dynamics.

A bright multisoliton solution of system (1) is obtained by using Hirota's direct method and the resulting $N$-soliton solution is expressed in the Gram determinant form. The nature of the soliton solution and their propagation is analyzed in detail for three different choices of nonlinearity coefficients: (i) positive nonlinearity coefficients ($c_{\ell}>0$), (ii) negative nonlinearity coefficients ($c_{\ell}<0$), and (iii) mixed-type nonlinearity coefficients with both positive and negative values for $c_{\ell}$. Additionally, we have investigated the soliton dynamics in the original 3-CNLS system, when LSRI takes place, by inverting the obtained solutions with appropriate inverse transformations and by constructing the asymptotic soliton of system (2). We show that the asymptotic soliton appearing in $q_1$ mode, when LSRI takes place, admits rich structure of profiles, like anti-dark, dark and gray solitons determined by the strength of XPM of the $q_1$ component with $q_2$ and $q_3$ components ($\sigma_{j1},~j=2,3$) for a given choice of nonlinearity coefficients $c_{\ell}$. However the $q_2$ and $q_3$ components support only bright asymptotic solitons irrespective of the strength of XPMs.

Different types of soliton collisions are explored and we identified that there exists only the overtaking collision of solitons in the $M$-YO  system (1). Particularly, the solitons appearing in the long-wave component undergo only an elastic collision accompanied by a phase-shift while the solitons in short-wave components undergo two types of energy-sharing collisions, namely, type-I and type-II energy-sharing collisions. In type-I energy-sharing collision, the solitons exhibit opposite kinds of energy switching for a given soliton between  the short-wave components, in which the energy in individual components and total energy are conserved. But in the case of type-II energy-sharing collisions, the switching nature of energy is similar in both short-wave components, thereby preserving the energy difference between the SW components. Apart from the energy switching behavior, these solitons also experience a phase-shift.  We have also investigated the soliton collisions in the 3-CNLS system, when there occurs a resonant interaction. Our study has revealed the fact that the solitons in the $q_1$ mode exhibit only elastic collisions for all types of soliton profiles (anti-dark, dark, and gray solitons) with a phase-shift. On the other hand, the solitons appearing in the $q_2$ and $q_3$ modes support only bright solitons and they undergo energy-sharing collision (both type-I and type-II energy-sharing collisions) and elastic collision. Also, here in contrary to the YO system, the solitons do undergo head-on collisions.

Additionally, we have studied the three-soliton collision briefly. The multisoliton collision takes place in a pairwise manner. The energy-sharing behavior and pair-wise nature of the three soliton collision lead to the possibility of restoring the state of a particular soliton after collision with two other solitons leaving them to undergo changes in their intensity. This property and the two different kinds of energy-sharing collisions make this $M$-YO system as a suitable candidate for performing logical operations in addition to the known Manakov system \cite{steig}. This suggests that the present study of the $M$-YO  system will find immediate applications in the context of soliton collision based optical computing and in realizing multistate logic. Apart from this, the type-II energy-sharing collision will find ramification in the direction of soliton amplification by their collision \cite{tkpre}, an interesting concept for achieving noiseless amplification of solitons in long distance optical communication without repeaters. One more application of the present study will be in the direction of pulse-shaping as evidenced from the different profiles of the asymptotic soliton in $q_1$ mode.

Mathematically, this study can be extended to construct bright-dark and dark-dark solitons and the underlying dynamics can be explored. Also, the higher dimensional version of the present general multicomponent Yajima-Oikawa system will be of future interest and work is in progress. As a next step, along this line, one can investigate the dynamics of multisoliton complexes, rogue waves and bound states in this new integrable multicomponent YO system, which may provide additional information. In view of these points, we do believe that our present study will be of broad interest and find multifaceted applications from both theoretical and experimental view points.

\section*{ACKNOWLEDGMENTS}
The work of T.K. is supported by the Department of Science and Technology, Government of India, in the form of a major research project. K.S. is grateful for the support of the Council of Scientific and Industrial Research, Government of India, with a Senior Research Fellowship. 

\appendix
\numberwithin{equation}{section}
\section{Arbitrary analysis for 3-component YO system}
In this appendix, we show the existence of a sufficient number of arbitrary parameters at each resonance value without any additional conditions for the completeness of the arbitrary analysis presented in Sec. \ref{secpain}. For simplification, we make use of the Kruskal ansatz \cite{weiss} for singular manifold $\phi(x,t)=x+\psi(t)$ and the coefficients $m_j^{(\ell)},~n_j^{(\ell)}$ and $l_j$ are chosen as functions of `$t$' alone. Here, we express the dependent variables up to the highest resonance value (i.e., $j=4$) as
\bes\bea
&&m^{(\ell)}=m_{0}^{(\ell)}\phi^{-1}+m_1^{(\ell)}+ m_2^{(\ell)} \phi+m_3^{(\ell)} \phi^2+m_4^{(\ell)} \phi^3,~\nonumber\\ &&n^{(\ell)}=n_{0}^{(\ell)}\phi^{-1}+n_1^{(\ell)}+ n_2^{(\ell)} \phi+n_3^{(\ell)} \phi^2+n_4^{(\ell)} \phi^3, \nonumber\\
&&l=l_{0}\phi^{-2}+l_1\phi^{-1}+ l_2+l_3 \phi+l_4 \phi^2,  \quad \ell=1,2,3.\nonumber
\eea\label{aa1}\ees
Then we substitute the above equations in Eq. (\ref{pa2}) and analyze the resulting set of equations arising at various powers of $\phi$ to show the existence of the required number of arbitrary parameters.

\noindent{\bf Coefficient of $\phi^{-3}$}:
At the order of $\phi^{-3}$, corresponding to the resonance $j=0$, we have only two equations ($l_0=-2$ and $\ds\sum_{\ell=1}^3 c_{\ell} m_0^{(\ell)} n_0^{(\ell)}=-2\psi_t$), which are nothing but the leading order equations (\ref{loe2}), for seven unknown parameters. Thus any five out of seven parameters ($m_0^{(\ell)},~n_0^{(\ell)}$, $\ell=1,2,3,$ and $l_0$) are arbitrary at $j=0$.

\noindent{\bf Coefficient of $\phi^{-2}$}:
We have the following equations at the coefficient of $\phi^{-2}$ for the resonance $j=1$,
\bes\bea
m_1^{(\ell)}&=&\frac{m_0^{(\ell)} }{2}\left(l_1 -{i \psi_t}\right),~
n_1^{(\ell)}=\frac{n_0^{(\ell)} }{2}\left(l_1 +{i \psi_t}\right),~~~\nonumber\\
l_1 \psi_t &=& \sum_{\ell=1}^3 c_{\ell} (m_1^{(\ell)} n_0^{(\ell)}+m_0^{(\ell)} n_1^{(\ell)}),  \; \ell=1,2,3. \nonumber
\eea\label{aaj1}\ees
From the above equations, we get $3\psi_t l_1=0 \Rightarrow l_1=0$. Hence we have seven equations for seven unknowns indicating that none of the parameters is arbitrary at this resonance, $j=1$.

\noindent{\bf Coefficient of $\phi^{-1}$}: At this order, we obtain $m_2^{(\ell)}=l_2 m_0^{(\ell)}+i m_{0t}^{(\ell)}$ and $n_2^{(\ell)}=l_2 n_0^{(\ell)}-i n_{0t}^{(\ell)}$, where $\ell=1,2,3$, i.e., six equations for seven unknown parameters. This shows the existence of one arbitrary parameter corresponding to the single resonance at $j=2$.

\noindent{\bf Coefficient of $\phi^{0}$}:
Here, we get $l_3=-\psi_{tt}/2$ and $\psi_t \psi_{tt}+l_{2t}+\ds\sum_{\ell=1}^3c_{\ell} (m_3^{(\ell)} n_0^{(\ell)}+m_0^{(\ell)} n_3^{(\ell)})=0$. Thus we have two equations for seven unknowns ($m_3^{(\ell)},~n_3^{(\ell)}$ and $l_3$, where $\ell=1,2,3$) which proves the arbitrariness of five parameters at the resonance $j=3$.\\
\noindent{\bf Coefficient of $\phi$}:
At this order, from (\ref{pa2}) we get
\bes\bea
&&4m_4^{(\ell)}+l_4m_0^{(\ell)}+2im_3^{(\ell)} \psi_t+\frac{i}{4}\psi_t \psi_{tt}m_0^{(\ell)} \nonumber\\ &&~~~~~~~+\frac{1}{2}(l_2^2m_0^{(\ell)}+2il_2 m_{0t}^{(\ell)}+il_{2t}m_0^{(\ell)}-m_{0tt}^{(\ell)})=0, \nonumber\\
&&4n_4^{(\ell)}+l_4n_0^{(\ell)}-2in_3^{(\ell)} \psi_t-\frac{i}{4}\psi_t \psi_{tt}n_0^{(\ell)} \nonumber\\&&~~~~~~+\frac{1}{2}(l_2^2n_0^{(\ell)}-2il_2 n_{0t}^{(\ell)}-il_{2t}n_0^{(\ell)}-n_{0tt}^{(\ell)})=0,
\nonumber\\
&& -2l_4 \psi_t+\frac{\psi_{ttt}}{2}+2\sum_{\ell=1}^3 c_{\ell} (m_4^{(\ell)} n_0^{(\ell)}+m_0^{(\ell)} n_4^{(\ell)})\nonumber\\&&~~~~~~+\frac{1}{2}\left[l_2^2\sum_{\ell=1}^3 c_{\ell} m_0^{(\ell)} n_0^{(\ell)}+\sum_{\ell=1}^3 c_{\ell} m_{0t}^{(\ell)} n_{0t}^{(\ell)}\right] \nonumber\\&&~~~~~~+\frac{il_2}{2}\left[\sum_{\ell=1}^3 c_{\ell} (m_{0t}^{(\ell)} n_0^{(\ell)}-m_0^{(\ell)} n_{0t}^{(\ell)})\right] \nonumber\\&&~~~~~~+i\psi_t\left[\sum_{\ell=1}^3c_{\ell} (m_3^{(\ell)} n_0^{(\ell)}-m_0^{(\ell)} n_3^{(\ell)})\right]=0, \nonumber
\eea\label{aaj4}\ees
where $\ell=1,2,3$. In the above, the former two equations can be made equivalent to the latter and hence we have only six equations for seven unknown parameters which proves the arbitrariness of one parameter at the resonance $j=4$. Thus we have shown the existence of a sufficient number of arbitrary parameters at each resonance value which proves the integrability of the three-component YO system in the Painlev\'e sense.
\section{Derivatives of Determinants $g^{(\ell)}$ and $f$}
The derivatives of the determinants $g^{(\ell)}$ and $f$ with respect to $x$ and $t$ used in obtaining the Jacobian identities (\ref{msolproof}) are given here:
\bea
g^{(\ell)}_{x}&=&
\left|
\begin{array}{cccc}
A & I & \phi & \phi_x \\
-I & B & {\bf 0}^T & {\bf 0}^T\\
{\bf 0} & a_{\ell} & 0&0\\
{\bf 0} & {\bf 0} & -1 & 0
\end{array}
\right|,\quad f_x= \left|
\begin{array}{ccc}
A & I & \phi \\
-I & B & {\bf 0}^T\\
-\phi^\dagger & {\bf 0} & 0
\end{array}
\right|,~~ \nonumber\\
f_{xx}&=&
\left|
\begin{array}{ccc}
A & I & \phi_x \\
-I & B & {\bf 0}^T\\
-\phi^{\dagger} & {\bf 0} & 0
\end{array}
\right|+
\left|
\begin{array}{ccc}
A & I & \phi \\
-I & B & {\bf 0}^T\\
-\phi^{\dagger}_x & {\bf 0} & 0
\end{array}
\right|,\nonumber\\
g^{(\ell)}_t &=& i
\left|
\begin{array}{cccc}
A & I & \phi &\phi_{xx}\\
-I & B & {\bf 0}^T& {\bf 0}^T\\
{\bf 0} & a_{\ell} & 0 &0\\
{\bf 0}& {\bf 0}&-1 &0
\end{array}
\right|+i
\left|
\begin{array}{cccc}
A & I & \phi &\phi_{x}\\
-I & B & {\bf 0}^T& {\bf 0}^T\\
{\bf 0} & a_{\ell} & 0 &0\\
-\phi^{\dagger}& {\bf 0}&0 &0
\end{array}
\right|,~~~~~~ \nonumber\eea \bea
g^{(\ell)}_{xx}&=&
\left|
\begin{array}{cccc}
A & I & \phi &\phi_{xx}\\
-I & B & {\bf 0}^T& {\bf 0}^T\\
{\bf 0} & a_{\ell} & 0 &0\\
{\bf 0}& {\bf 0}&-1 &0
\end{array}
\right|-
\left|
\begin{array}{cccc}
A & I & \phi &\phi_{x}\\
-I & B & {\bf 0}^T& {\bf 0}^T\\
{\bf 0} & a_{\ell} & 0 &0\\
-\phi^{\dagger}& {\bf 0}&0 &0
\end{array}
\right|,~~\nonumber\\
f_{t}&=&i
\left|
\begin{array}{ccc}
A & I & \phi_x \\
-I & B & {\bf 0}^T\\
-\phi^{\dagger} & {\bf 0} & 0
\end{array}
\right|-i
\left|
\begin{array}{ccc}
A & I & \phi \\
-I & B & {\bf 0}^T\\
-\phi^{\dagger}_x & {\bf 0} & 0
\end{array}
\right|.\nonumber
\eea
The complex conjugate of $g^{(\ell)}$, $f_t$ and $f_{xt}$ can be written as \newline $\qquad g^{(\ell)*}=
-\left|
\begin{array}{ccc}
A & I & {\bf 0}^T\\
-I & B & -a_{\ell}^{\dagger}\\
-\phi^\dagger & {\bf 0} & 0
\end{array}
\right|$, \newline $\qquad f_t=\sum_{\ell=1}^M {c_{\ell}}
\left|
\begin{array}{ccc}
A & I & {\bf 0}^T \\
-I & B & -a_{\ell}^{\dagger}\\
{\bf 0} & a_{\ell} & 0
\end{array}
\right|$, \newline and \newline $\qquad f_{xt}=\sum_{\ell=1}^M {c_{\ell}}
\left|
\begin{array}{cccc}
A & I & \phi &{\bf 0}^T \\
-I & B & {\bf 0}^T& -a_{\ell}^{\dagger}\\
-\phi^{\dagger}&{\bf 0} & 0 & 0 \\
{\bf 0}& a_{\ell}&0 &0
\end{array}
\right|$.\newline


\end{document}